\documentclass[11pt]{article}
\usepackage{graphicx}
\usepackage{rotating}

\usepackage{indentfirst}

\newcommand{\BABARPubYear}    {03}

\newcommand{\BABARConfNumber} {001}
\newcommand{\SLACPubNumber} {9676}

\input pubboard/babarsym

\newcommand{\onreslumi}  {\mbox{56.4 \invfb}}
\newcommand{\offreslumi} {\mbox{6.4 \invfb}}
\newcommand{\bbpairs}    {\mbox{61.6 million}}
\newcommand{\nbb}    {\mbox{N$_{\BB}$}}
\newcommand{\fz}           {\mbox{$f_0(980)$}}

\newcommand{\ppK}          {\mbox{$K^+ \pi^- \pi^+$}}

\newcommand{\Kstarpi}      {\mbox{$\Kstarz(892) \pi^+$}}
\newcommand{\BpmKstarpi}   {\mbox{$B^+ \to \Kstarpi$}}
\newcommand{\KstarIIpi}    {\mbox{$\Kstarz(1430) \pi^+$}}

\newcommand{\RhoK}         {\mbox{$\rho^0(770) K^+$}}
\newcommand{\BpmRhoK}      {\mbox{$B^+ \to \RhoK$}}
\newcommand{\fzK}          {\mbox{$\fz K^+$}}
\newcommand{\BpmfzK}       {\mbox{$B^+ \to \fzK$}}
\newcommand{\fzII}        {\mbox{$f_2(1270)$}}
\newcommand{\fzIIK}        {\mbox{$f_2(1270) K^+$}}

\newcommand{\ChiczK}       {\mbox{$\chi_{c0} K^+$}}
\newcommand{\BpmChiczK}    {\mbox{$B^+ \to \ChiczK$}}
\newcommand{\Dzpi}         {\mbox{$\Dzb \pi^+$}}
\newcommand{\BpmDzpi}      {\mbox{$B^+ \to \Dzpi$}}

\newcommand{\BrDzpi}{184.6 \pm 3.2 \pm 9.7}

\newcommand{\BrKstarpi}{10.3\pm1.2^{+1.0}_{-2.7}}

\newcommand{\BrKstarIIpi}{25.1\pm2.0^{+11.0}_{-5.7}}

\newcommand{\BrfzK} {9.2\pm1.2^{+2.1}_{-2.6}}

\newcommand{\BrChiczK}{1.46\pm0.35\pm0.12}

\newcommand{\BrKstarpiVal}  {(\BrKstarpi)\times 10^{-6}}
\newcommand{\BrKstarIIpiVal}{(\BrKstarIIpi)\times 10^{-6}}

\newcommand{\BrfzKVal}      {(\BrfzK)\times 10^{-6}}

\newcommand{\BrChiczKVal}   {(\BrChiczK)\times 10^{-6}}
\newcommand{\BrDzpiVal}     {(\BrDzpi)\times 10^{-6}}

\newcommand{\DE}{\ensuremath{\Delta E}}

\newcommand{\calB}{\mbox{${\cal B}$}}

\newcommand{\xf}{\mbox{${\cal F}$}}

\newcommand{\half}{\mbox{${1\over2}$}}
\newcommand{\pvec}{{\bf p}}

\def\Y#1S{{\Upsilon\rm(#1S)}}

\def\ra{\rightarrow}

\def\beq{\begin{equation}}
\def\eeq{\end{equation}}

\long\def\inst#1{\par\nobreak\kern 4pt\nobreak
    {\it #1}\par\vskip 10pt plus 3pt minus 3pt}

\graphicspath{{Figures/}}

\begin{document}
{\pagestyle{empty}

\begin{flushright}
\babar-CONF-\BABARPubYear/\BABARConfNumber \\
SLAC-PUB-\SLACPubNumber \\
March 2003 \\
\end{flushright}

\par\vskip 3cm
\begin{center}
\Large \bf Measurements of the Branching Fractions of Charged \B\ Decays to \ppK\ Final States
\end{center}
\bigskip

\begin{center}
\large The \babar\ Collaboration\\
\mbox{ }\\
\today
\end{center}
\bigskip \bigskip

\begin{center}
\large \bf Abstract
\end{center}

We present preliminary results of searches for exclusive charged--$B$
decays to $K^{\pm}\pi^{\mp}\pi^{\pm}$ from \bbpairs\ \BB\ pairs
collected at the $\FourS$ resonance with the \babar\ detector at the
SLAC \pep2\ asymmetric \B\ Factory. The Dalitz plot is divided into
eight regions and, using a maximum--likelihood fit, we measure
statistically significant yields in all regions.  We interpret the
results as the following branching fractions averaged over
charged--conjugate states: $\calB(\BpmKstarpi, \Kstarz \ra K^+\pi^-) =
\BrKstarpiVal$, $\calB(\BpmfzK, f_0\ra \pi^+\pi^-) = \BrfzKVal$,
$\calB(\BpmChiczK, \chi_{c0}\ra \pi^+\pi^-) = \BrChiczKVal$ and
$\calB(\BpmDzpi, \Dzb \ra K^+\pi^-) = \BrDzpiVal$.  The first
uncertainty is statistical and the second is systematic and includes
resonance--model and interference uncertainties.  We give 90\% confidence-level 
upper limits on the branching fractions of the following channels:
$\calB(\BpmRhoK) < 6.2\times 10^{-6}$ and $\calB(B^+ \to K^+ \pi^- \pi^+$
non--resonant) $< 17\times 10^{-6}$.

\vfill
\begin{center}
Presented at the XVII$^{th}$ Rencontres de la Vall\'ee d'Aoste, \\
3/9---3/15/2003, La Thuile, Vall\'ee d'Aoste, Italy
\end{center}

\vspace{1.0cm}
\begin{center}
{\em Stanford Linear Accelerator Center, Stanford University, 
Stanford, CA 94309} \\ \vspace{0.1cm}\hrule\vspace{0.1cm}
Work supported in part by Department of Energy contract DE-AC03-76SF00515.
\end{center}

\newpage
} 

\begin{center}
\small

The \babar\ Collaboration,
\bigskip

%
B.~Aubert,
R.~Barate,
D.~Boutigny,
J.-M.~Gaillard,
A.~Hicheur,
Y.~Karyotakis,
J.~P.~Lees,
P.~Robbe,
V.~Tisserand,
A.~Zghiche
\inst{Laboratoire de Physique des Particules, F-74941 Annecy-le-Vieux, France }
% ...

\end{center}\newpage

\setcounter{footnote}{0}

 \section{Introduction}
\label{sec:Introduction}

The study of charmless hadronic \B\ decays can make important
contributions to the understanding of \CP\ violation in the Standard
Model as well as to models of hadronic decays. The measurement of $B^+$
\footnote{Charge--conjugate states are included throughout this document.} 
decays to the final state $K^+\pi^-\pi^+$ via
intermediate resonances can be used to search for direct \CP\ violation.
The three--body final state is unique in the search for weak phases since
it is possible to isolate the strong phase variation for overlapping 
resonances.  There has been recent theoretical progress on proposed methods for extracting the
Cabibbo--Kobayashi--Maskawa angle $\gamma$ through the interference of
\ChiczK\ with other \ppK\ final states \cite{gamma1,ref:GammaPaper}.
Study of these decays can also help clarify the nature of the resonances
involved, not all of which are well understood. We present preliminary
results on the branching fractions of $B^+$ decays to the final state
\ppK\ both non--resonant and by way of intermediate resonances.

 \section{The \babar\ Detector and Dataset}
\label{sec:babar}

The data used in this analysis were collected with the \babar\ detector
at the \pep2\ asymmetric--energy \epem\ storage ring at SLAC.  The data sample
consists of \bbpairs\ \BB\ pairs, corresponding to an integrated
luminosity of \onreslumi\ collected on the $\FourS$ resonance (10.58
\gev) during the 2000-2001 run.  In addition, a total integrated
luminosity of \offreslumi\ was taken 40~\mev\ below the $\FourS$
resonance, and was used to study backgrounds from continuum production.

The \babar\ detector is described in detail elsewhere~\cite{ref:babar}; 
the main parts relevant for this analysis are the tracking and particle identification sub--systems.

The 5--layer double--sided silicon vertex tracker (SVT) measures the
impact parameters, angles, and transverse momenta of tracks.  Outside
the SVT is a 40--layer drift chamber (DCH), which measures the transverse
momenta of tracks from their curvature in the 1.5-T solenoidal magnetic
field.  The ionization energy loss of charged tracks, $\dedx$, in the
SVT and DCH is used in the particle--type identification.  The tracking
system has a momentum resolution of 0.5\% for a transverse momentum of
1.0~\gevc\ and a typical \dedx\ resolution of $7.5$\%.

Surrounding the DCH is a detector of internally--reflected Cherenkov
radiation (DIRC), which provides charged--hadron identification in the
barrel region.  
The separation between pions and kaons varies from $>
8\sigma$ at 2.0\gevc to 2.5$\sigma$ at 4.0\gevc, where $\sigma$ is
the average resolution on the Cherenkov angle.  

The DIRC is surrounded by a Cesium Iodide electromagnetic calorimeter
 (EMC), which is used to measure the energies and
angular positions of photons and electrons with excellent resolution.
In this analysis, the EMC is used to veto electrons.

\section{Analysis Method}
\label{sec:Analysis}

The \ppK\ final state can be represented in a Dalitz plot \cite{Dalitz}.
The many resonant $B$ decay modes form bands in such a plot.  These
resonances often overlap and interfere so the whole Dalitz plot should
be considered before assigning a branching fraction to a specific mode.
This analysis divides the Dalitz plot into regions, each of which is expected 
to be dominated by a particular contribution.  First the yields in
these regions are determined, using a maximum--likelihood fit, with no
assumption on the form of the intermediate resonances.  We then interpret
these yields as branching fractions, assuming a model for the
contributions to the Dalitz plot.  We also consider the uncertainty of
the model and the effect of overlap and interference between these
contributions.

\subsection{Dalitz Plot Regions}

\begin{figure}[tbh!]
\begin{center}
\includegraphics[width=10cm]{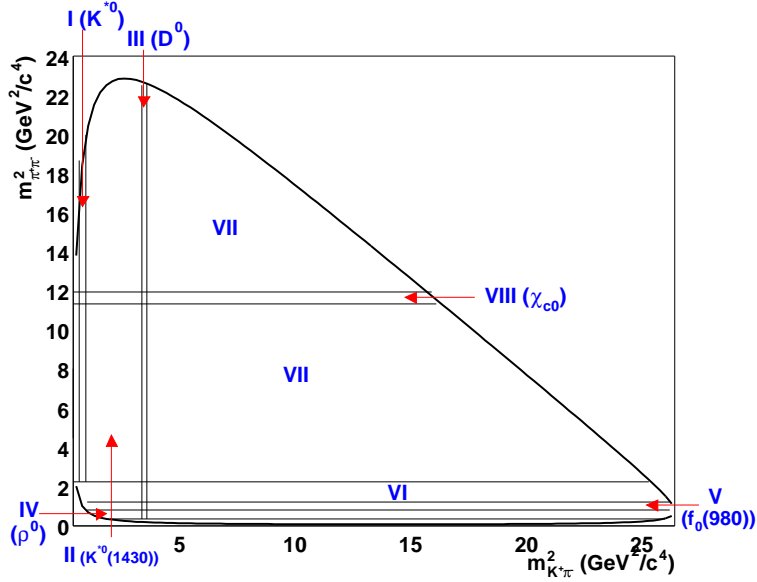}
\caption{An illustration of the different regions in the Dalitz plot 
as defined in Table \ref{tab:regions}.}
\label{fig:regions}
\end{center}
\end{figure}

The \ppK\ Dalitz plot is divided into eight regions. 
Each region is designed to contain a large proportion of the decays of the 
expected dominant resonance (if any) and to
minimize contributions from neighboring modes. 
The definitions of the regions are given in
Table \ref{tab:regions} and they are illustrated in
Figure \ref{fig:regions}.

Regions I, II and III are characterized by narrow bands in the invariant
mass of the $K^+\pi^-$ system, $m_{K\pi}$.  Region I is expected to be 
dominated by the
$K^{*0}(892)$.  The primary resonance contribution to region II, labeled
``higher $K^{*0}$'', is not currently known.
The areas where these bands cross the $\pi\pi$ resonances ($\rho
(770), \fz$, ``higher $f$'' and $\chi_{c0}$) are excluded to limit
biases from interference.  Region III is dominated by the production
of $\Dzb \pi^+$. The relatively high branching fraction for this mode
allows it to be used to correct for differences between data and
simulated events and evaluate systematic uncertainties.

Regions IV, V and VI are characterized by narrow bands in the $\pi^+\pi^-$ invariant mass, $m_{\pi\pi}$.
Regions IV and V are expected to be dominated by the $\rho (770)$ and \fz,  
respectively. 
The resonance contributions to region VI (``higher $f$'') are not well
defined. 
The area where these regions would intersect the \Dzb\ band, $1.8<m_{K\pi}<1.9\gevcc$, is excluded from regions IV, V and VI.
The area where the other $K\pi$ resonances cross is not excluded from regions IV, V 
and VI as the overall interference uncertainty 
on $\calB(\BpmfzK)$ and $\calB(\BpmRhoK)$ was estimated to be smaller 
when this area is excluded.  
Region VII, denoted ``high mass'', could contain higher charmless and charmonium
resonances as well as a non--resonant contribution.

Region VIII is dominated by \ChiczK.
A lower limit on $m_{K\pi}$ ensures that this region is free of
contamination from resonances in regions I, II and III.  The area
$3.355<m_{\pi\pi}<3.475\gevcc$ is removed from all other regions to
avoid this charmonium background.

\begin{table}[htb]
\caption{Regions of the \ppK\ Dalitz plot. The regions are
kinematically defined by the $m_{K\pi}$ and $m_{\pi\pi}$ mass selection
criteria. The symbol ``$!$'' is used to imply exclusion. The assumed dominant contribution 
is noted for each region.}
\begin{center}
\begin{tabular}{|l|l||c|c|}
\hline
Region & Dominant & \multicolumn{2}{|c|}{Selection Criteria }\\ \cline{3-4}
       & Contribution & $m_{K \pi}$ (\gevcc) & $m_{\pi\pi}$ (\gevcc) \\ \hline \hline
I    & \Kstarpi   & $0.816<m_{K\pi}<0.976$ & $m_{\pi\pi}>1.5$ \ $!(3.355<m_{\pi\pi}<3.475)$ \\
II   & ``higher $K^{*0}$'' $\pi^+$& $0.976<m_{K\pi}<1.8$ & $m_{\pi\pi}>1.5$ \ $!(3.355<m_{\pi\pi}<3.475)$  \\
III  & \Dzpi      & $1.835 < m_{K \pi} < 1.895$ &  $!(3.355<m_{\pi\pi}<3.475)$ \\
IV   & \RhoK      & $!(1.8<m_{K\pi}<1.9)$  & $0.6 < m_{\pi\pi} < 0.9$   \\
V    & \fzK       & $!(1.8<m_{K\pi}<1.9)$  &  $0.9  < m_{\pi\pi} < 1.1 $\\
VI   & ``higher $f$'' $K^+$& $!(1.8<m_{K\pi}<1.9)$  &  $1.1  < m_{\pi\pi} < 1.5 $\\
VII  & ``high mass''  & $m_{K \pi} >  1.9$ & $m_{\pi\pi}>1.5$ \ $!(3.355<m_{\pi\pi}<3.475)$ \\ 
VIII & \ChiczK    & $m_{K \pi} >  1.9$ & $3.37<m_{\pi\pi}<3.46$ \\ \hline
\end{tabular}
\label{tab:regions}
\end{center}
\end{table}

\subsection{Candidate Selection}
\label{sec:Selection}

\B\ candidates are reconstructed from charged tracks that have at least 12
hits in the DCH, a maximum momentum of 10~\gevc, a minimum transverse
momentum of 100~\mevc, and originate from the beam--spot. The \B\ candidates
are formed from three--charged--track combinations and particle 
identification selections are applied. Mass hypotheses are assigned
accordingly and the \B\ candidates' energies and momenta are required to
satisfy appropriate kinematic constraints.

Two kinematic variables are defined that are included in the
maximum--likelihood fit described later. The first of these is the
beam--energy--substituted mass $\mes = \sqrt{(E^2_b - \pvec^2_B)}$.  The
energy of the \B\ candidate is defined as $E_b = (\half s + \pvec_0
\cdot \pvec_B)/E_0$, where $\sqrt{s}$ and $E_0$ are the total energies
of the \epem\ system in the center--of--mass (CM) and laboratory frames,
respectively, and $\pvec_0$ and $\pvec_B$ are the three--momenta in the
laboratory frame of the \epem\ system and the \B\ candidate,
respectively. The $\mes$ value should be close to the nominal \B\ mass
for signal events.

The second variable used is the energy difference, $\DE$, between the
energy in the CM of the reconstructed \B\ candidate, $E_B^*$, and the
beam energy, $\DE = E_B^* - \sqrt{s}/2$.  \DE\ is dependent on the mass
hypotheses of the tracks. To each track a mass is assigned appropriate
for the particle identification selections applied.
For signal events, \DE\ should be centered at zero.

To identify charged pions and kaons, we use \dedx information from the
SVT and DCH for tracks with momenta below 700~\mevc, 
the number of photons measured by the DIRC for tracks with momenta
above 500~\mevc, and the Cherenkov angle for tracks with momenta
above 700~\mevc.  Kaons are selected with requirements on the product of
likelihood ratios determined from these measurements and pions are
required to fail the kaon selection.  The average selection
efficiency for kaons in our final state that have passed the tracking
requirements is $\sim 80\%$ including geometrical acceptance, while the
misidentification probability of pions as kaons is below 5\% at all
momenta.  The kaon veto on pions in our final state is $\sim 98\%$
efficient.  We veto electron candidates by requiring that they fail a
selection based on information from \dedx, shower shapes in the EMC and
the ratio of the shower energy and track momentum.  The probability of
misidentifying electrons as pions is approximately 5\%, while the
probability of misidentifying pions as electrons is $\sim 0.2\%$.

\subsection{Background Suppression and Characterization}
\label{sec:suppression}

The dominant background in this analysis is from light quark and charm
continuum production. This background is suppressed by imposing
requirements on topological event shape variables, and the remainder is
characterized and parameterized in the maximum--likelihood fits used to
extract the signal yield from the data.  There are also backgrounds from
$B$ decays which, although contributing far fewer events, can be
more difficult to separate from the signal and must also be
parameterized in the fit, which is described in detail later.

The event shape variables used to suppress continuum background are calculated in the $\FourS$ rest frame.
The first is the cosine of the angle $\theta_T$ between the thrust axis
of the selected $B$ candidate and the thrust
axis of the rest of the event, {\em i.e.}, of all charged tracks and neutral
particles not in the $B$ meson candidate. For continuum backgrounds,
the directions of the two axes tend to be aligned because the
daughters of the reconstructed candidate generally lie along the dijet
axis of such events so the distribution of $\cos\theta_T$
is strongly peaked towards $\pm 1$. For $B$ events, however, the
distribution is isotropic because the decay products from the two 
$B$ mesons are independent of each other and the \B\ 
mesons have very low momenta in the $\FourS$ rest frame. 
To improve the signal--to--background ratio, the criterion
 $|\cos\theta_T| < 0.9$ is applied to all regions except
region III (\Dzb). This selection removes about 60\% of the
continuum background while retaining over 90\% of the signal. The
$\cos\theta_T$ requirement in the \Dzb\ region is varied to estimate 
a systematic uncertainty on the efficiency of this criterion in the 
other regions.

We also make use of a Fisher discriminant~\cite{Fisher}, 
using a linear combination of the angle between the \B\ candidate
momentum and the beam direction; the angle between the \B\ candidate
thrust axis and the beam direction; and the energy flow of the rest of
the event into each of 9 independent concentric $10^{\circ}$ cones around the
thrust axis of the reconstructed $B$ \cite{CLEOCones}.
The variables and weights of the Fisher discriminant were chosen to
optimize the separation of our final state and the continuum background
after the $\cos\theta_T$ criterion has been applied.  The resulting
Fisher variable, ${\cal{F}}$, is used in the maximum--likelihood fit.

The $B$ decay backgrounds are from four main sources: combinatorial
background from three unrelated tracks; three-- and four-body $B\ra D$
decays; charmless four--body decays with a missing track and
three--body decays with one or more particles misidentified.  These
backgrounds are reduced by the particle identification selections and,
where possible, removed by vetoing regions of the invariant--mass
spectra of pairs of the final--state particles.  The influence of
remaining specific backgrounds on the signal yield obtained from the
maximum--likelihood fit was established using test fits with
Monte--Carlo simulated data (MC) with the expected number of signal,
continuum background and $B$ background events.
Background modes that significantly contributed to the signal 
yields in these tests are parameterized for the final fit to the data.
However, if the background contributes only a few events, it is instead
subtracted from the signal yield.

The combinatorial background from $B$ decays is less than 2\% of the
continuum background. The shape of its \DE\ and \mes\ distributions are
similar to the continuum background and these events were found to be
fitted as such in the test fits so no additional parameterisation was
required.

The particle--misidentification background has several sources.  $B^+
\to \jpsi K^+$ and $B^+ \to \Psi (2S) K^+$ decays which contribute
through muon/pion misidentification are removed completely by
excluding events with $2.97<m_{\pi\pi}<3.17\gevcc$ or
$3.56<m_{\pi\pi}<3.76\gevcc$.  This also excludes other \B\ backgrounds 
containing \jpsi\ or $\Psi (2S)$, such as  $B^+ \to \jpsi K^{*+}$.
The contributions from $B^+\ra K^+e^+e^-$
are negligible due to the electron veto.  Non--resonant $B^+\ra
K^+\mu^+\mu^-$ decays contribute with $18\pm 7$ events uniformly
distributed over the Dalitz plot, using the branching fraction from
\cite{KllConf}. Events are expected only in regions II, VI and VII where
they are subtracted from the yields.  Contributions from the $B^+\ra
\pi^+\pi^-\pi^+$ and $B^+\ra K^+K^-\pi^+$ final states may affect
regions I, II, IV, VI and VII.  The channels concerned have not yet been
observed and so 
an additional negative uncertainty is added to the signal yield equal to
the number of events expected in the fit corresponding to the upper
limit measured in \cite{JohnConf}.  This is 20 events for region VII and
6 events or fewer for all other regions.

The decay $\BpmDzpi, \Dzb \ra \Kp\pim\piz$ contributes to the measured
signal yield when not parameterized in the test fits in regions II and
VII, while $B^+ \rightarrow \Dzb \rho^+(770)$ with $\Dzb \rightarrow K^+
\pi^-$ and $\rho^+ \rightarrow \pi^+ \pi^0$ has a significant effect on
the test fits in region VII.  These modes are therefore parameterized in
the final fit for the affected regions.
 
The decay $B^+ \rightarrow \eta' K^+ $ with $\eta' \rightarrow \rho^0
(770) \gamma$, $ \rho^0 \to \pi^+ \pi^-$ is the only charmless channel
with a four--body final state found to contribute significantly. The
expected number of events is 31 and 12 in regions IV and V, respectively,
using the branching ratio measured in \cite{etapPRL}. If not
parameterized, these contribute to the signal yields in the test fits.
This mode is therefore included in the final fit for those regions.

\subsection{The Maximum--Likelihood Fit}
\label{sec:FitVariables}

We form Probability Density Functions (PDFs) for the 3 variables \mes,
\DE\ and $\mathcal{F}$ in each region. For each hypothesis $l$ (signal,
continuum background, and if applicable, $B$ background) these three
PDFs form a product $P_{l}$, which models that particular hypothesis.
These products are functions of the variables $\vec{x}$ and parameters
$\vec{\alpha}$ of the PDFs.  The likelihood for an event $j$ is formed
by summing the products over the $M$ hypotheses, with each product
weighted by the number of events in that hypothesis $n_l$.  A product
over the $N$ events in the data sample of the per--event likelihoods
along with a Poisson factor forms the total likelihood function
$\mathcal{L}$, written in equation~(\ref{eq:Likelihood}). 

\begin{equation}
  \label{eq:Likelihood}
  \mbox{$\mathcal{L}$} \,=\, \exp\left(-\sum_{i=1}^{M} n_i\right)\, \prod_{j=1}^N 
\,\left(\sum_{l=1}^M n_{l} \, P_{l}(\vec{\alpha},\vec{x_j})\right).
\end{equation}
 
The fit is
performed in two stages for each region. First, one--dimensional fits
are performed on the particular data samples detailed below in order to
determine the PDF parameters. Then the multivariate fit is performed on
the final data samples to extract the signal and continuum background
yields.

The signal PDF parameters, particularly the width of the \DE\ 
distribution, vary across the Dalitz plane. Therefore, they are found
for each region separately using Monte--Carlo--simulated signal of the
expected dominant resonance where available and, otherwise, non--resonant \ppK\ 
selected for that region.
Some differences have been observed between MC and data in the distributions 
of the  \mes, \DE\ and $\mathcal{F}$ variables.
These differences are measured in the high--statistics  $B^+\ra\Dzb\pip$ 
dominated region and are used to correct all regions where all the signal 
PDF parameters are fixed in the final fits. 
The \mes\ PDF is a Crystal Ball function\cite{CBall}, the \DE\ PDF 
is two Gaussians with equal
means and the $\cal{F}$ PDF is a ``bifurcated'' Gaussian (a Gaussian 
with different widths above and below the mean).

The background PDF parameters are also found to vary across the Dalitz
plane and so are determined individually for each region.  The $\cal{F}$
PDF is a bifurcated Gaussian where the parameters are determined using
data sidebands defined by $0.1 < |\DE | <0.35 \gev$ except in
regions II and VII where, due to $B$ background, only the positive \DE\ 
sideband is used.  The \mes\ variable is parameterized by the Argus threshold
function~\cite{ref:Argus} with two parameters: a fixed kinematic
endpoint and a shape parameter which is left to float
in the final fits.  The \DE\ PDF is a first--order polynomial where the
gradient is also left to float in the final fits.

 \section{Physics Results}
\label{sec:Physics}

\subsection{Fit Results}

The signal yields for the various regions of the \ppK\ Dalitz plot are
shown in Table \ref{table:SigYields}. The first uncertainty is
statistical and the second is systematic. The systematic uncertainty
is from the uncertainties on the PDF parameters and from the $B$
background subtraction.  The yields are found to be statistically
significant in all regions ($>5\sigma$), where the statistical
significance is taken as $\sqrt{ -2 \log
(\mathcal{L}_{max}/\mathcal{L}_{(n_{sig}=0)} )}$.

\begin{table}[htb]
\caption{Signal yields for the regions of the Dalitz plot before and
  after \B\ background subtraction. The first uncertainty is statistical, the
second systematic.}
\begin{center}
\begin{tabular}{|l||c|c|c|}
\hline
Region  & Signal Yield & Signal Yield after \B\\
        &              & background subtraction \\
    \hline \hline
I    & $161 \pm 18 \pm  4$ & $161 \pm 18 \pm 4$ \\
II   & $406 \pm 28 \pm 12$ & $405 \pm 28 ^{+12}_{-13}$\\
III  & $3755 \pm 66 \pm 11$ & $3755 \pm 66 \pm 11$\\
IV   & $ 66 \pm 15 \pm  3$ & $ 66 \pm 15 ^{+3}_{-7}$ \\
V    & $179 \pm 19 \pm  5$ & $179 \pm 19 \pm 5$ \\
VI   & $127 \pm 19 \pm  5$ & $126 \pm 19 \pm 5$ \\
VII  & $147 \pm 23 \pm  7$ & $133 \pm 23 ^{+9}_{-22}$ \\
VIII & $ 26 \pm  6 \pm  1$ & $26 \pm 6 \pm 1$ \\ \hline
\end{tabular}
\label{table:SigYields}
\end{center}
\end{table}

\begin{figure}[p]
  \centering
\resizebox{\textwidth}{!}{
\begin{tabular}{ccc}
    \includegraphics{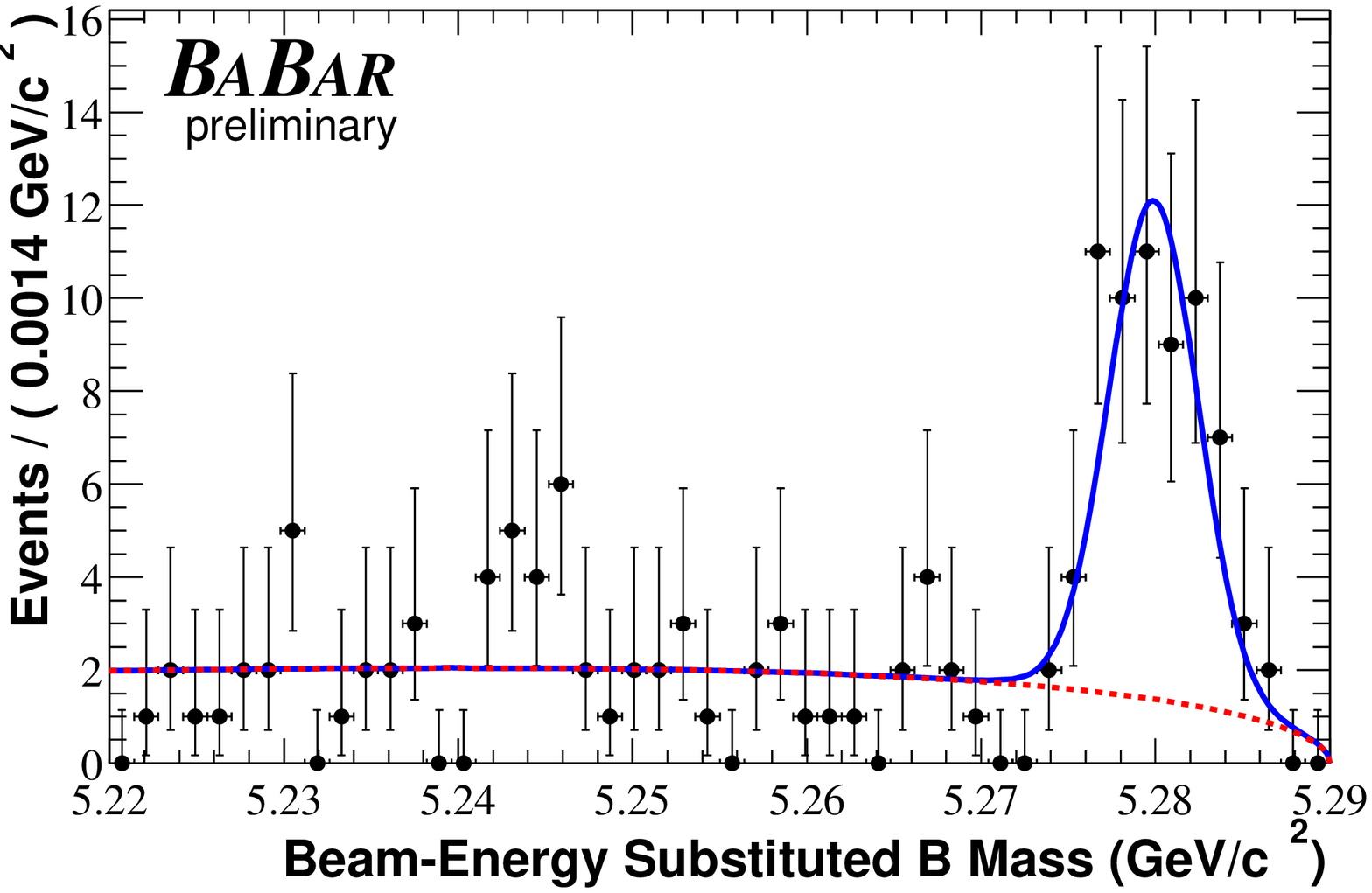} &
    \includegraphics{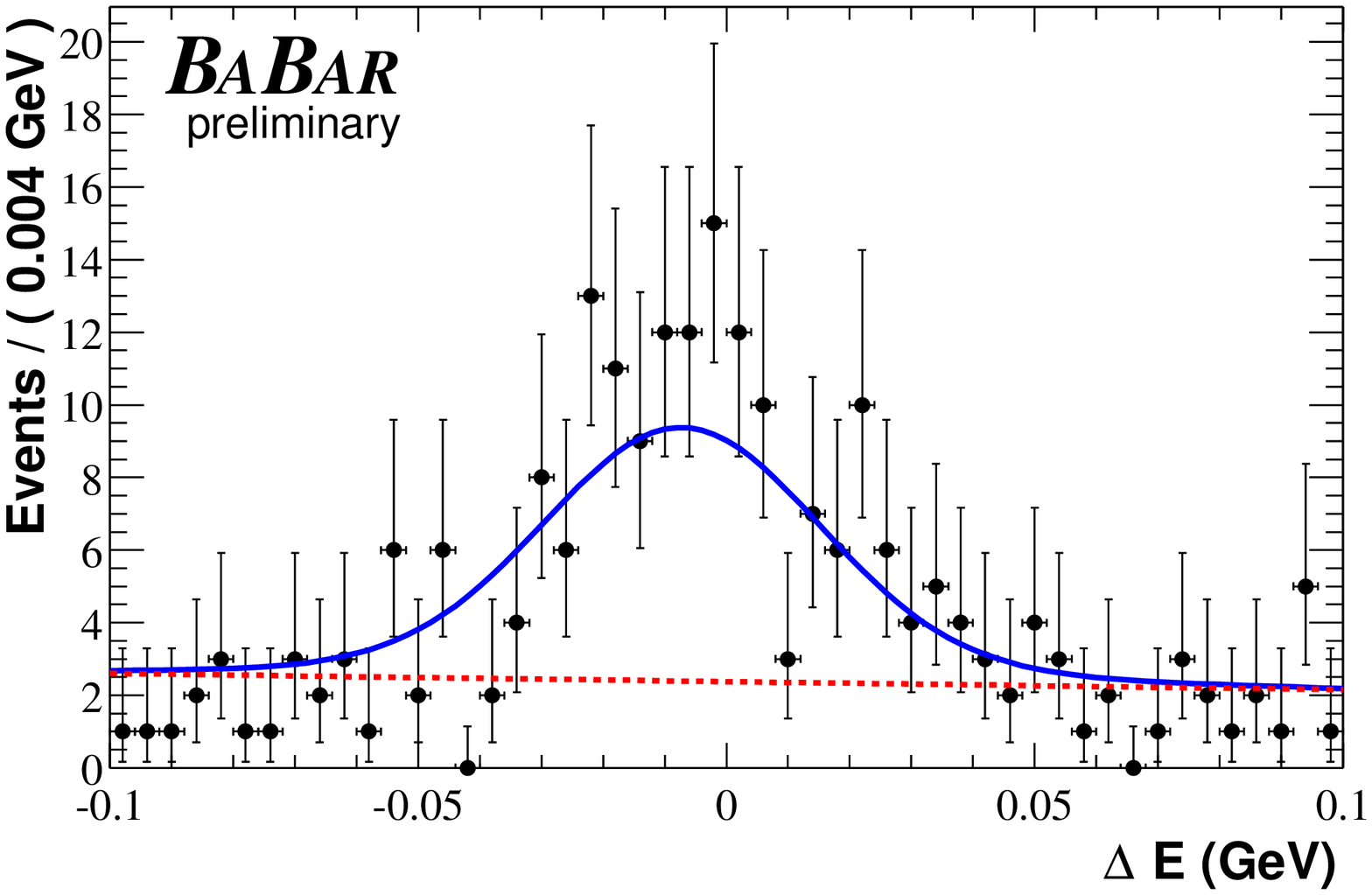} &
    \includegraphics{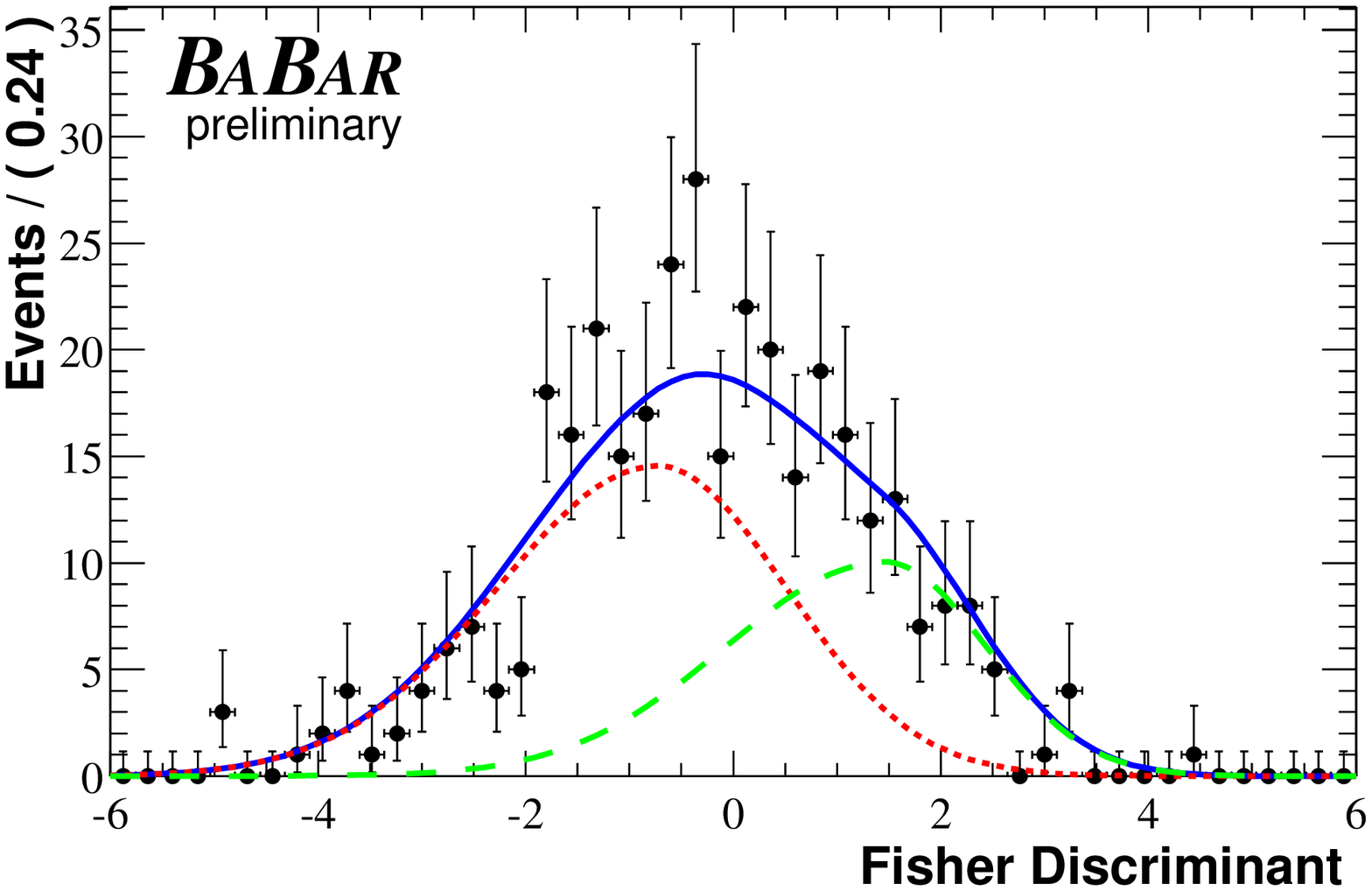} \\
    \includegraphics{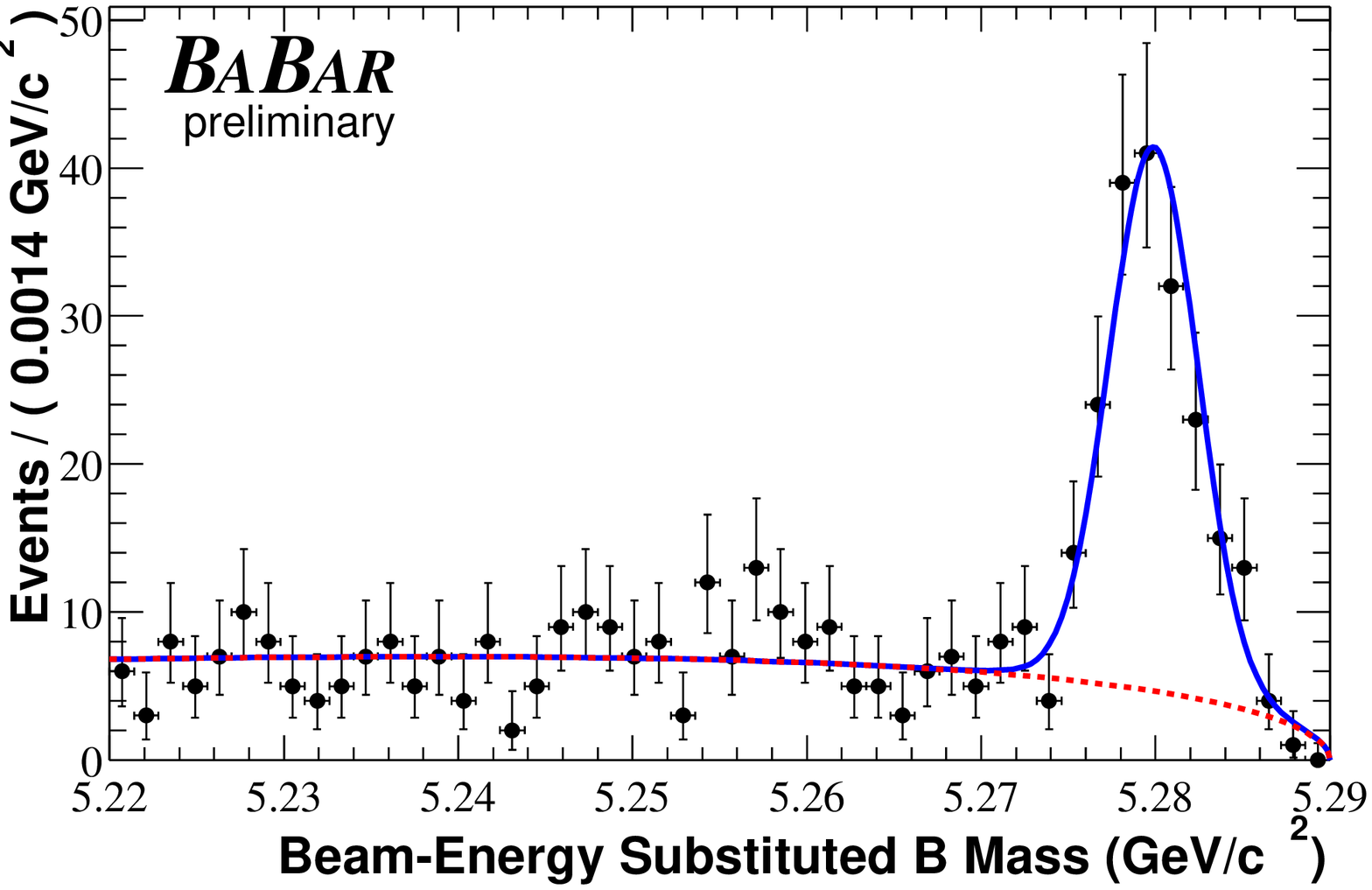} &
    \includegraphics{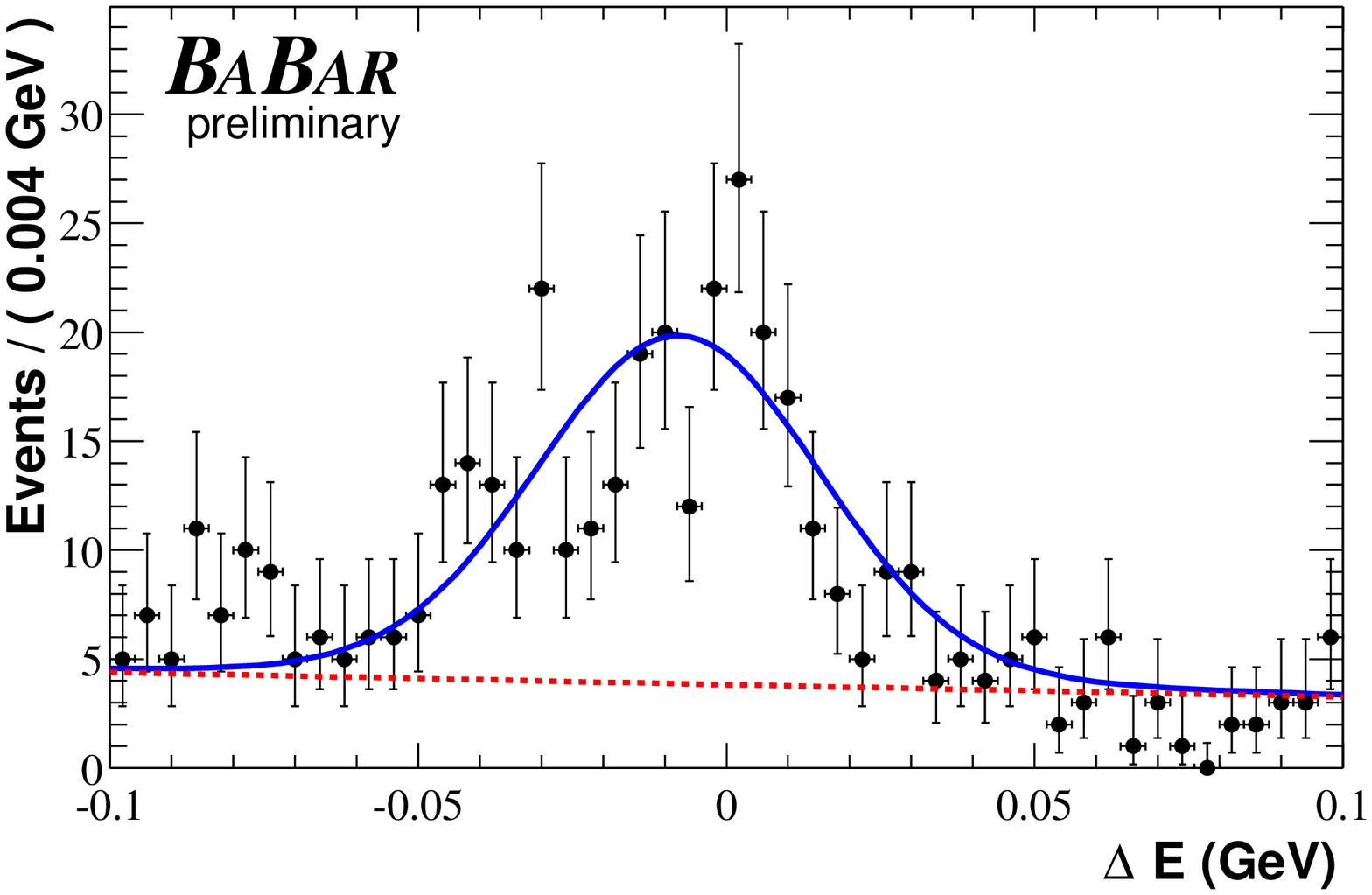} &
    \includegraphics{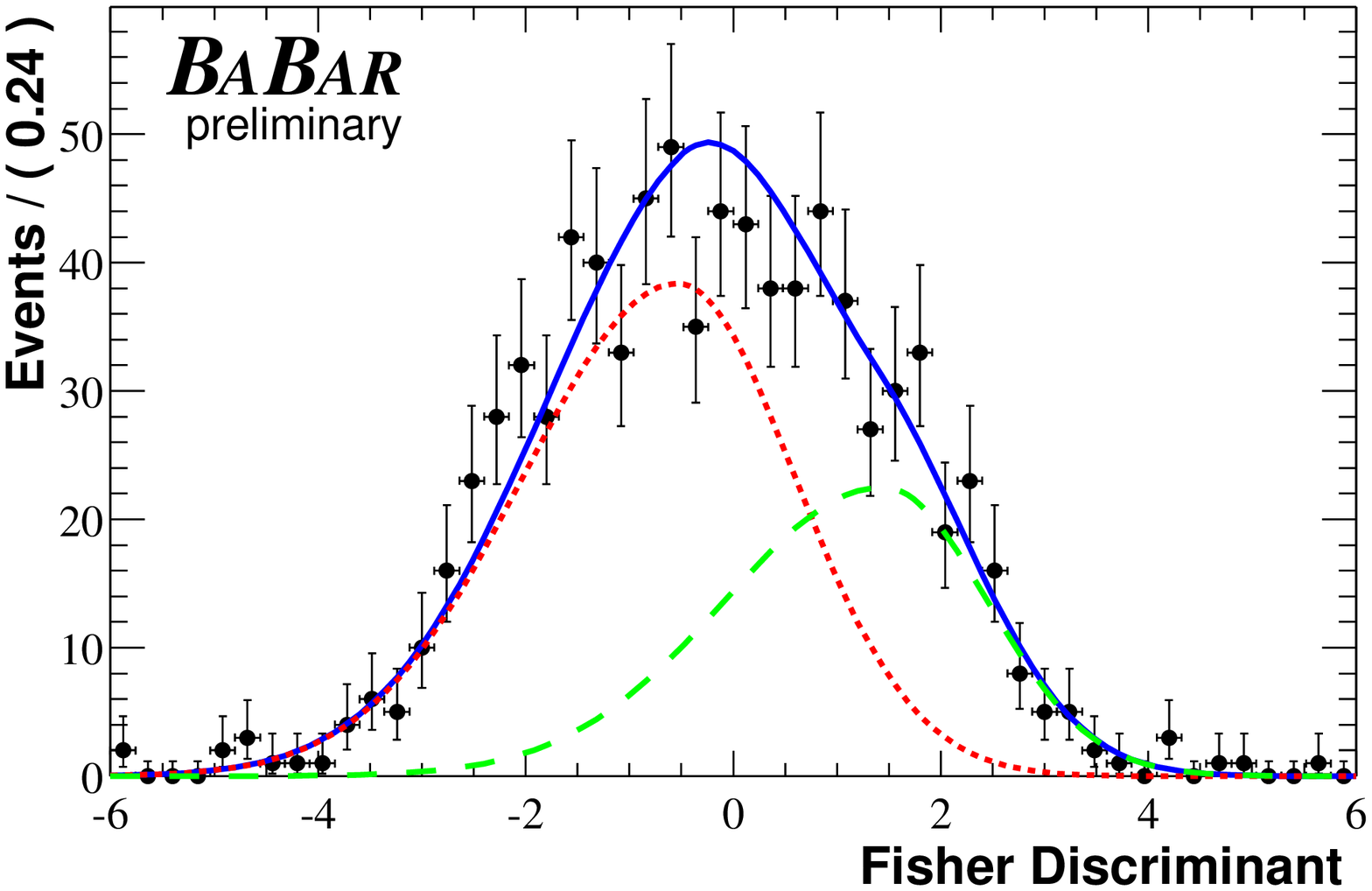} \\
    \includegraphics{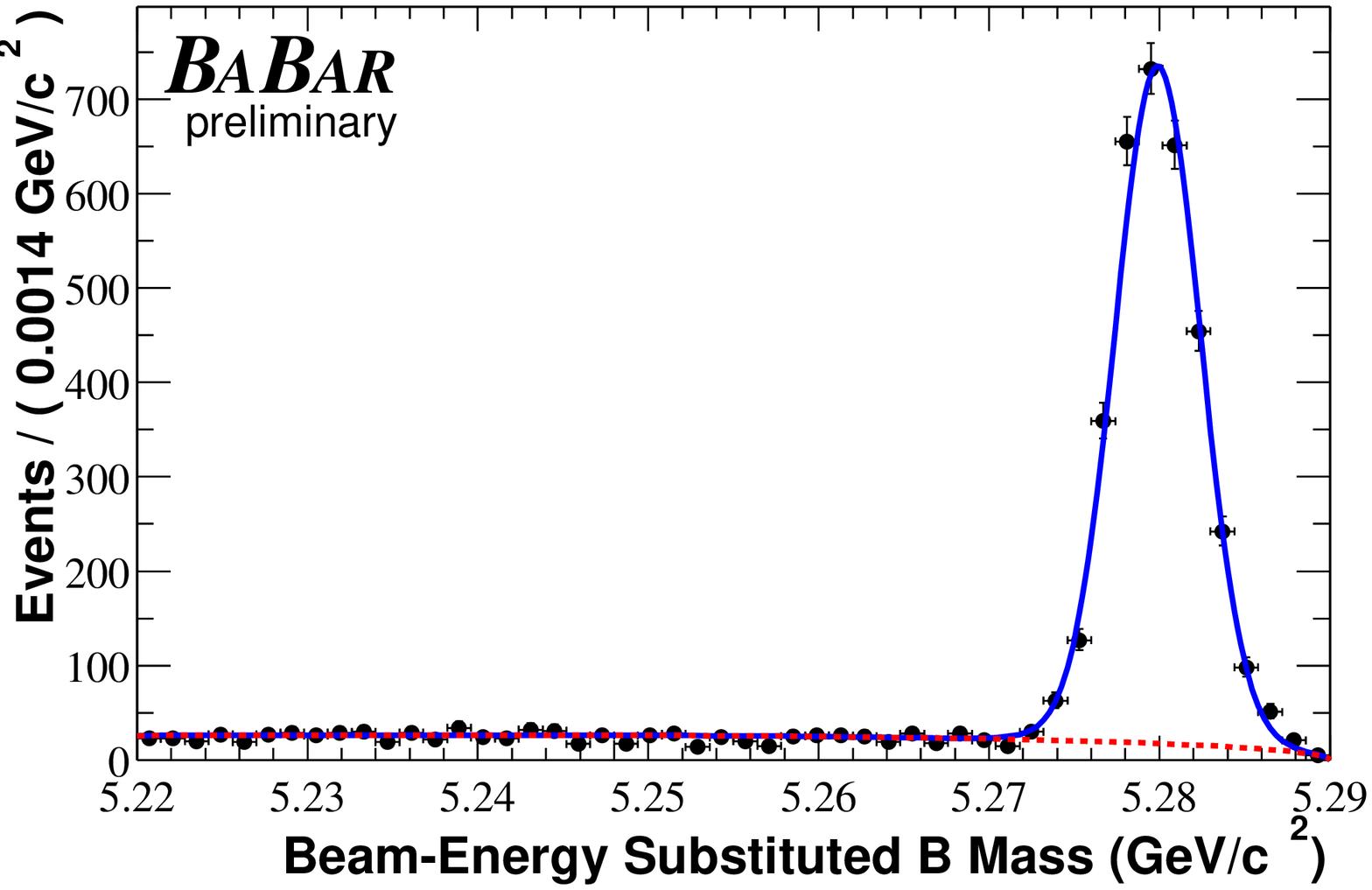} &
    \includegraphics{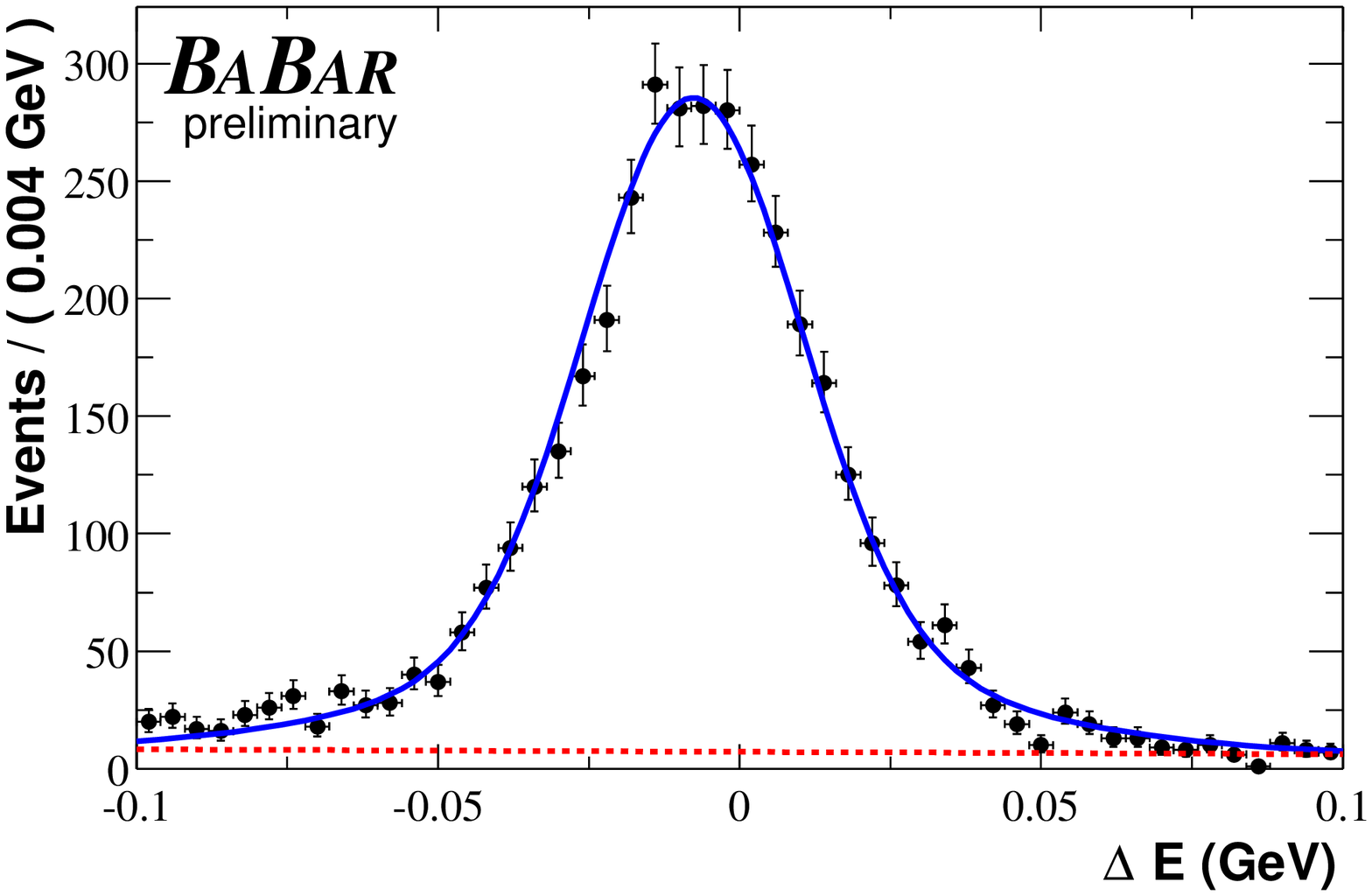} &
    \includegraphics{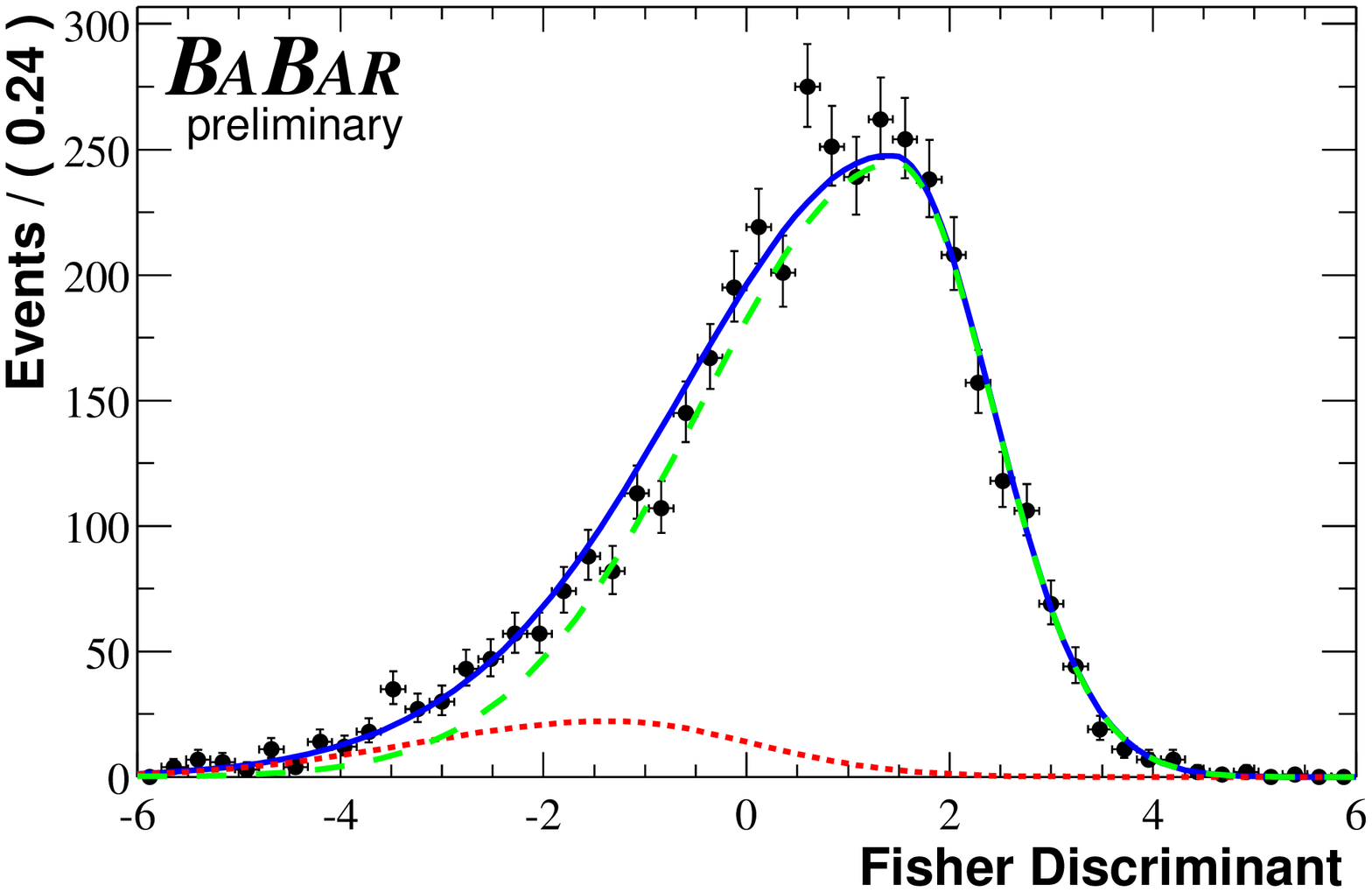} \\
    \includegraphics{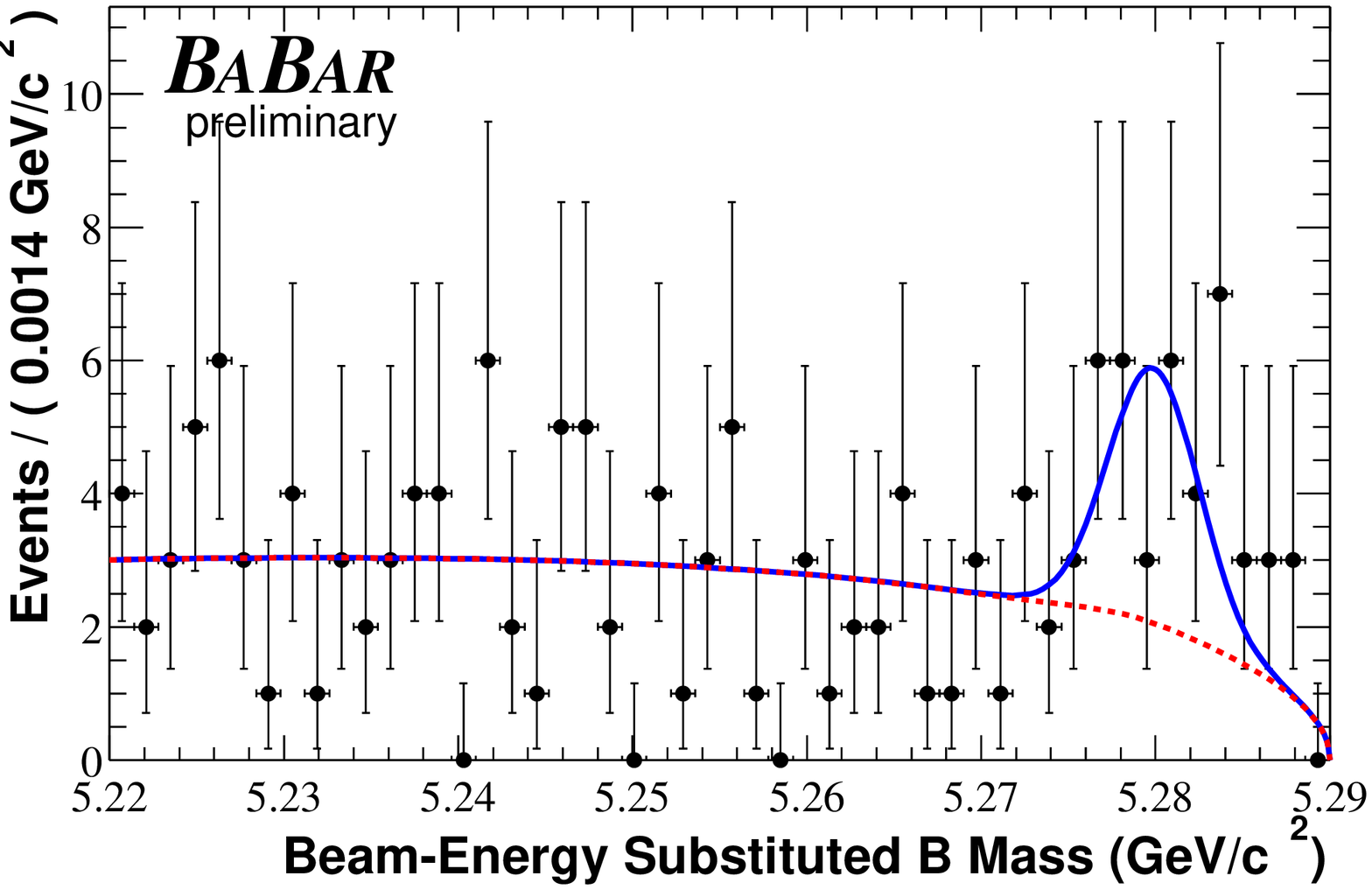} &
    \includegraphics{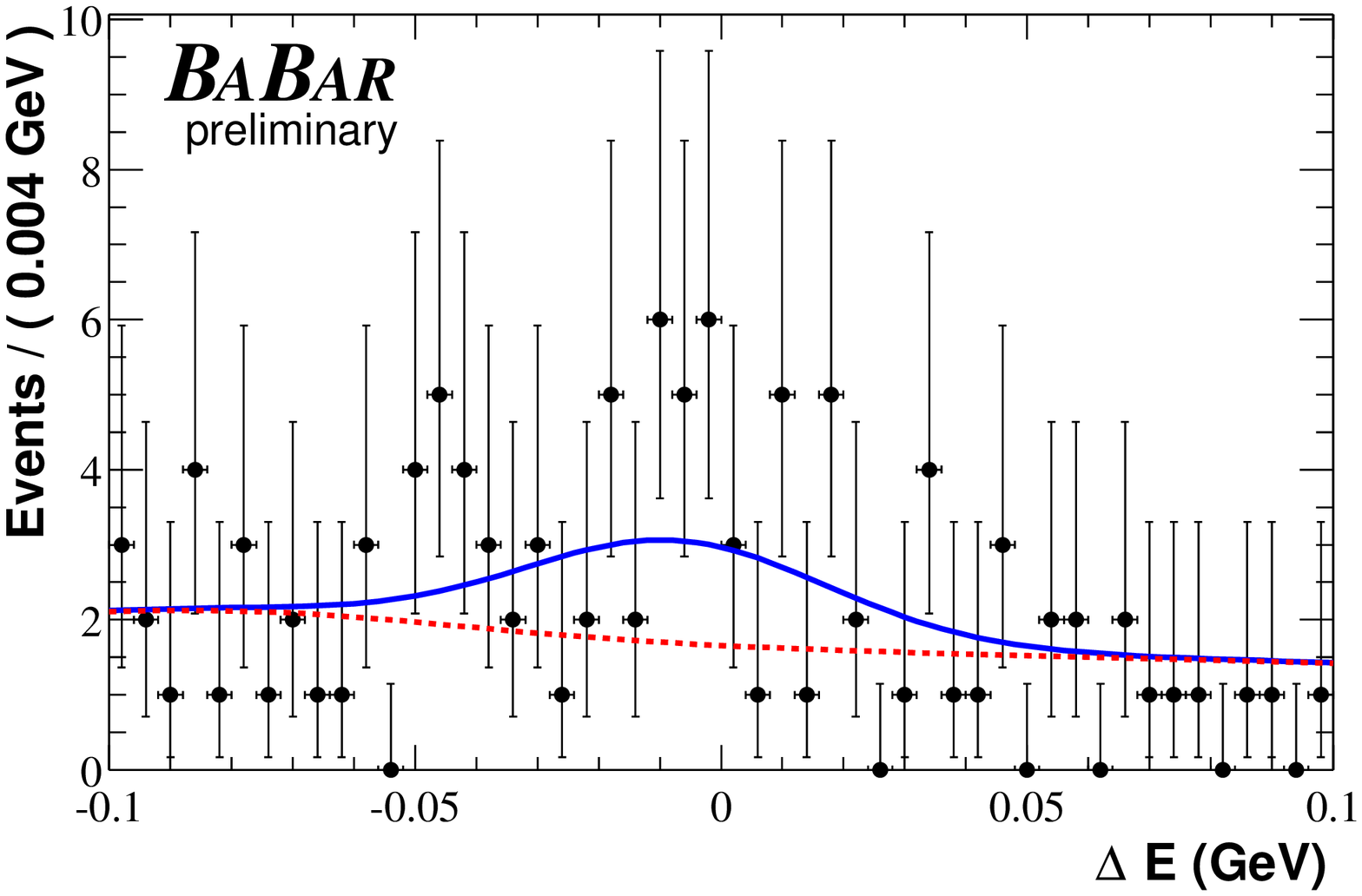} &
    \includegraphics{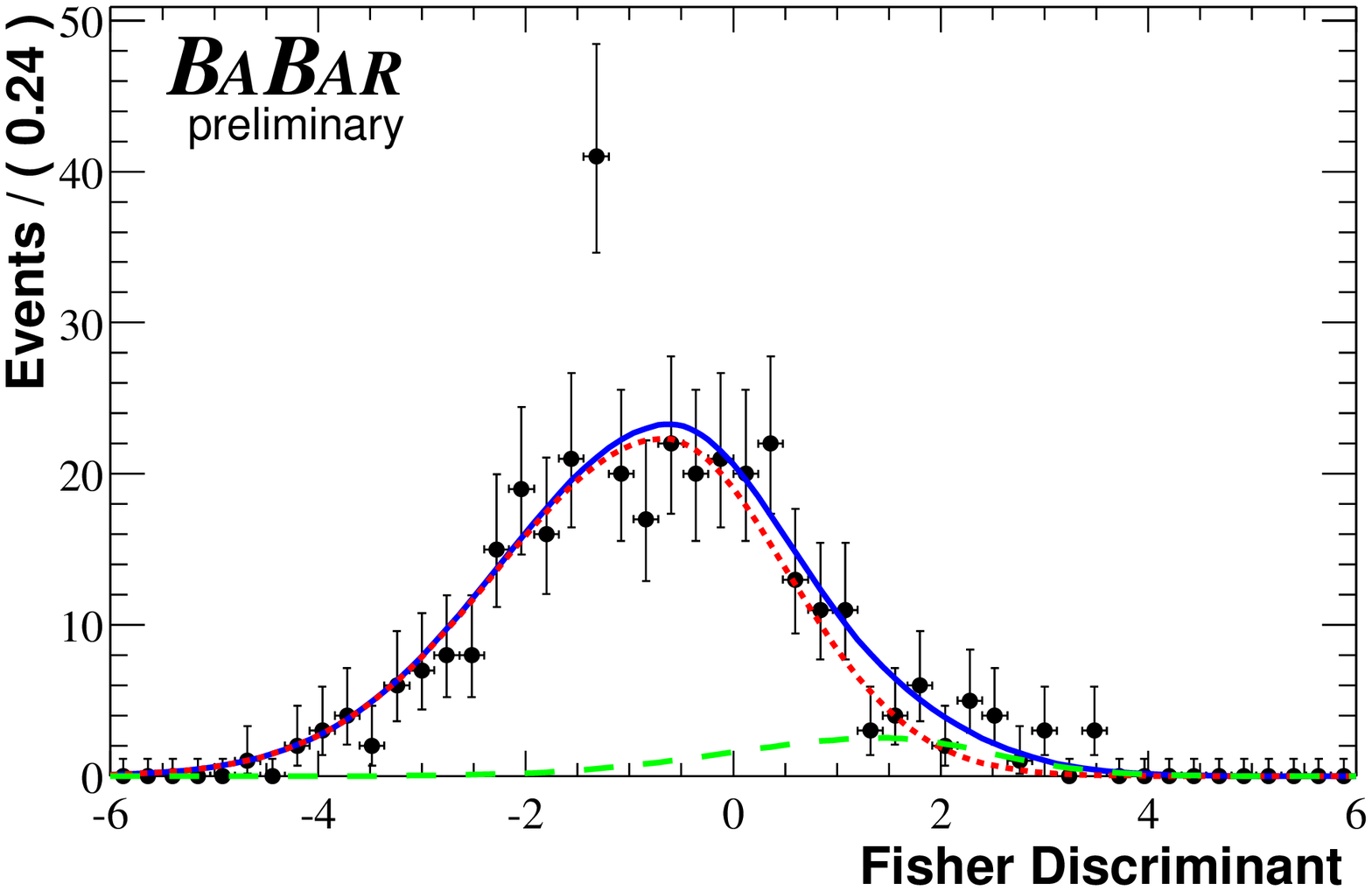}
\end{tabular}
}
\caption{Projection plots in \mes, \DE \ and \xf, produced by selecting on the event likelihood ratio 
  formed from the other two fit variables for (from top to bottom)
  regions I, II, III and IV. The superimposed curve is a projection of
  the full fit with the background component shown as a dotted line
  and, for \xf, the signal component shown as a dashed
  line.}
\label{fig:ProjPlots1}  
\end{figure}

\begin{figure}[p]
  \centering
\resizebox{\textwidth}{!}{
\begin{tabular}{ccc}
    \includegraphics{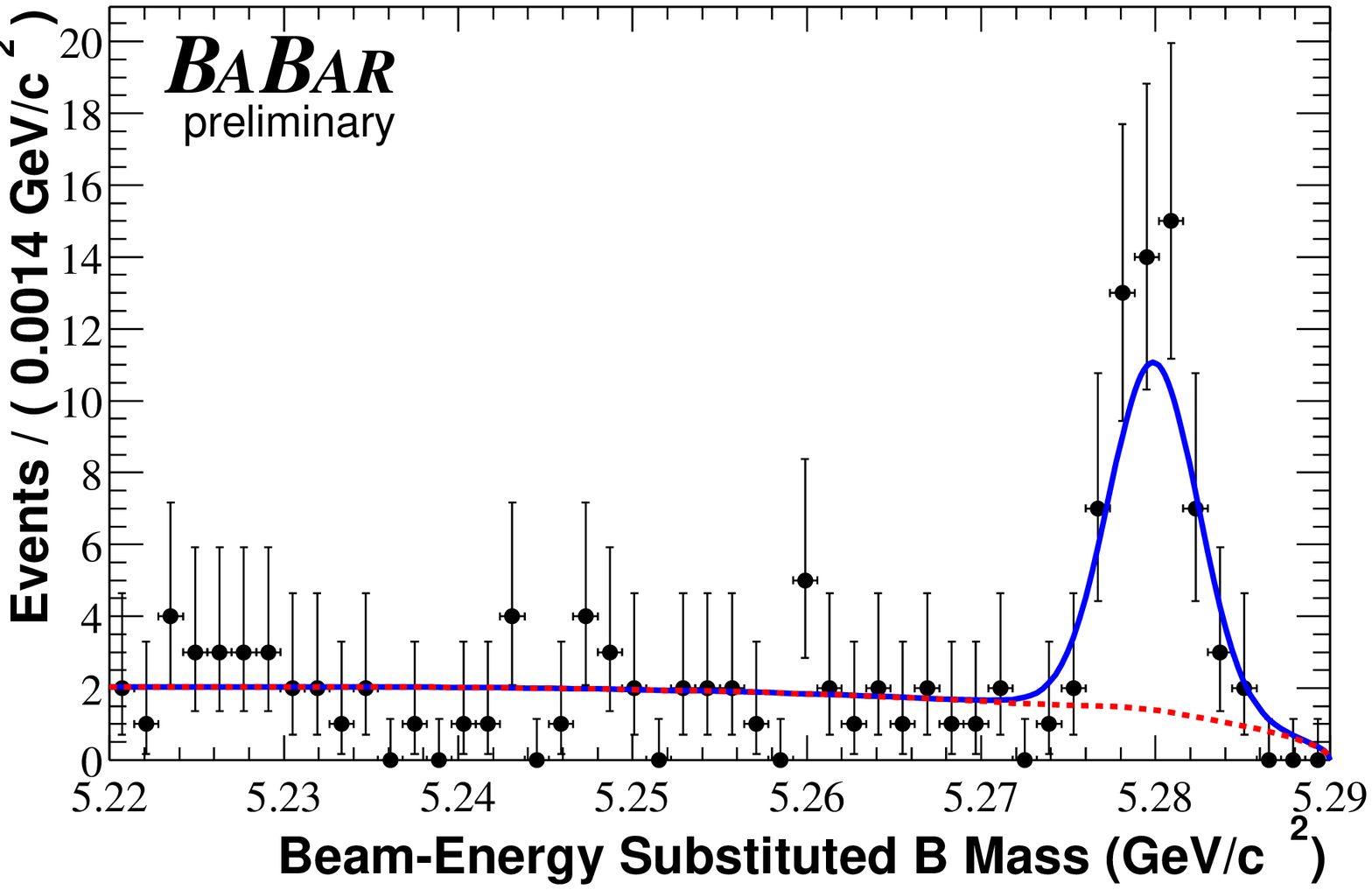} &
    \includegraphics{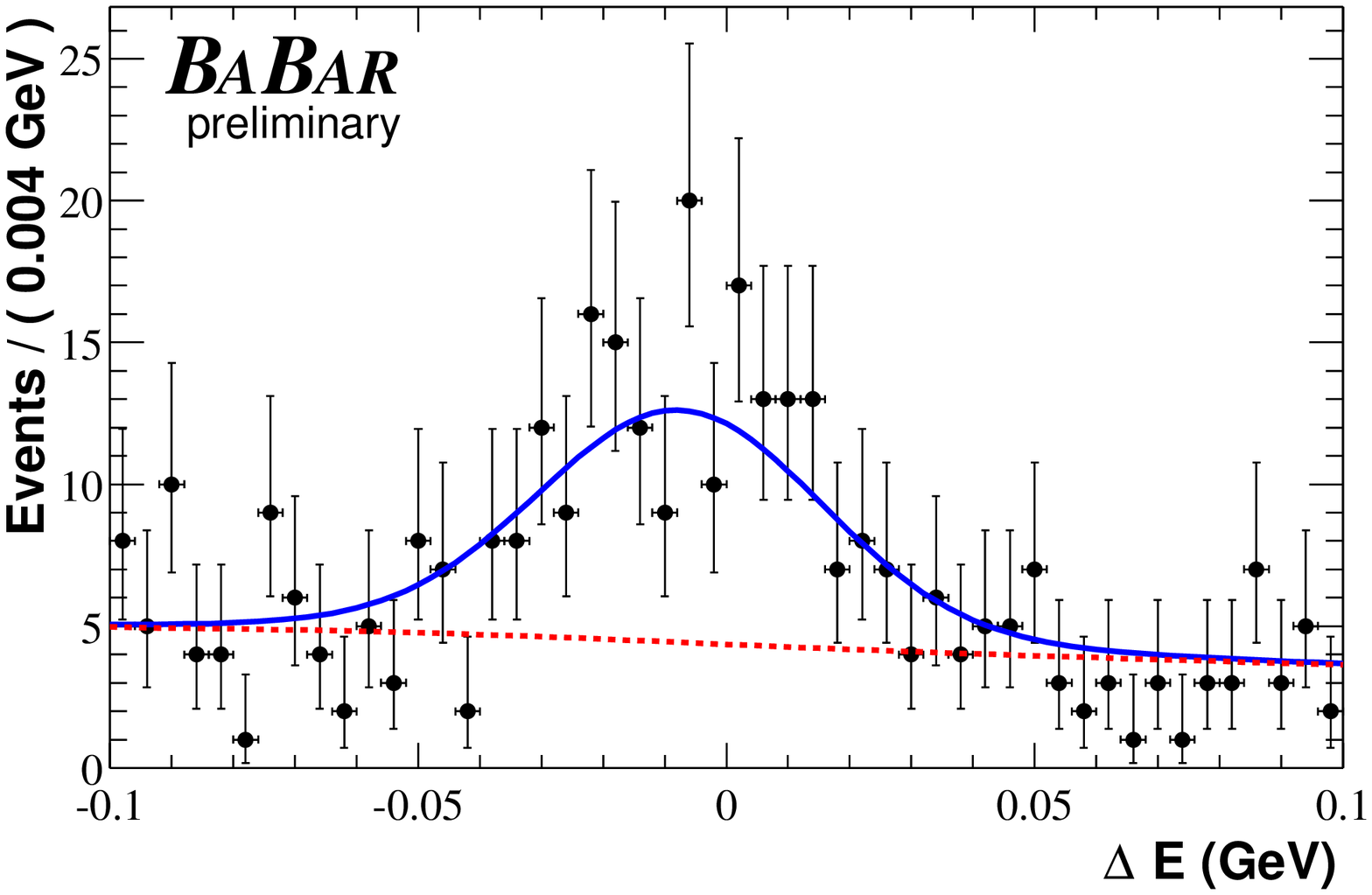} &
    \includegraphics{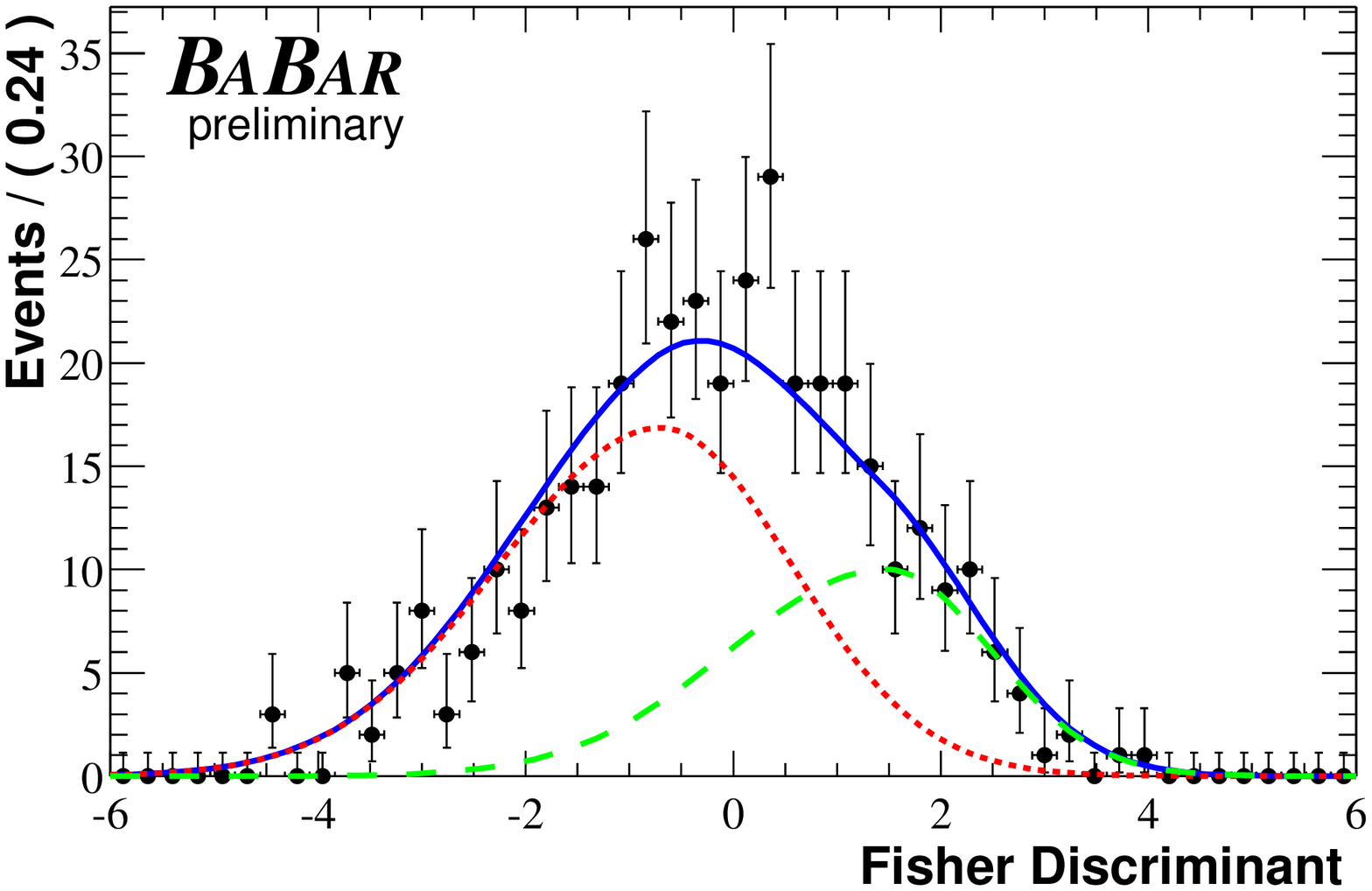} \\
    \includegraphics{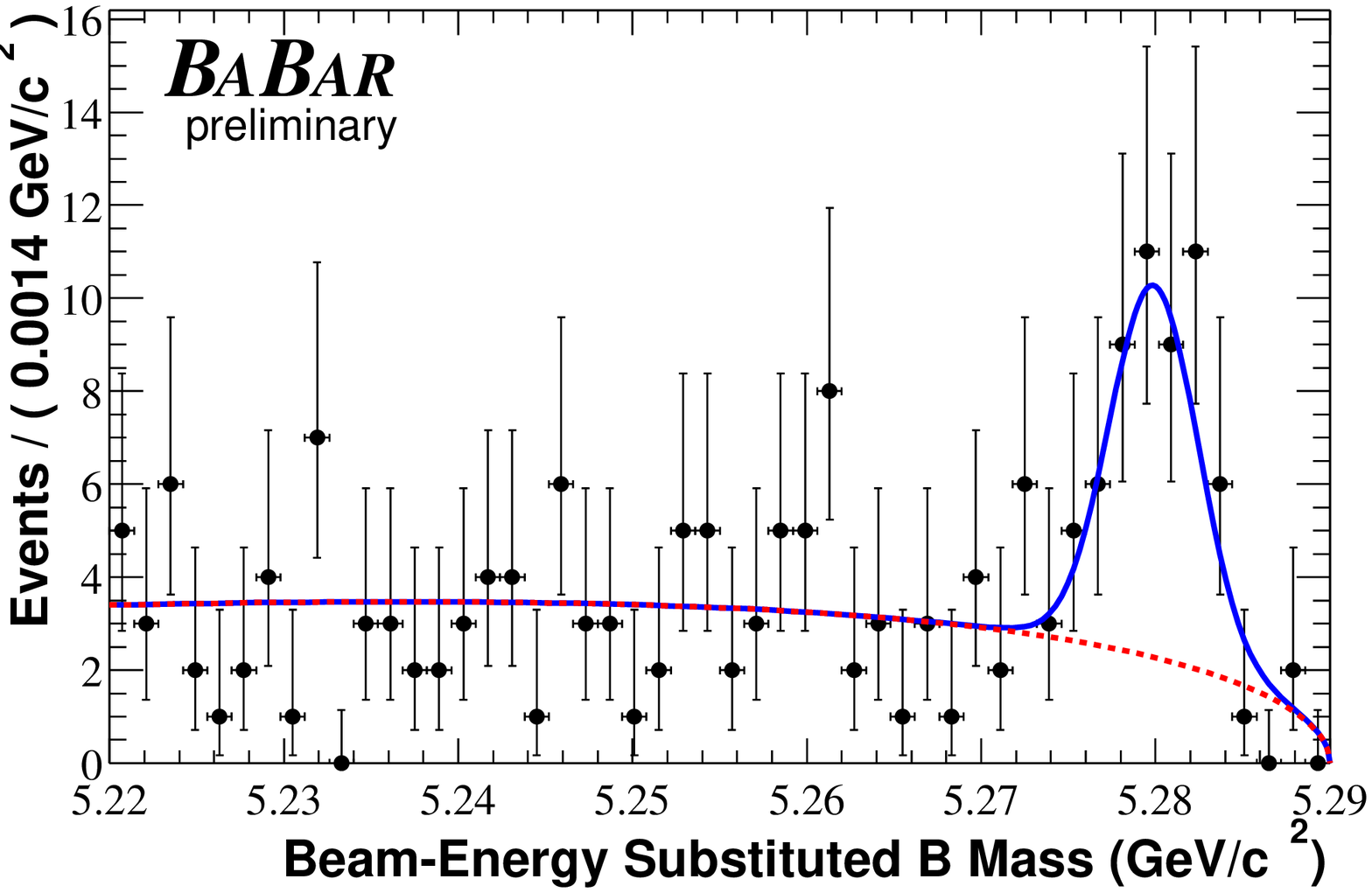} &
    \includegraphics{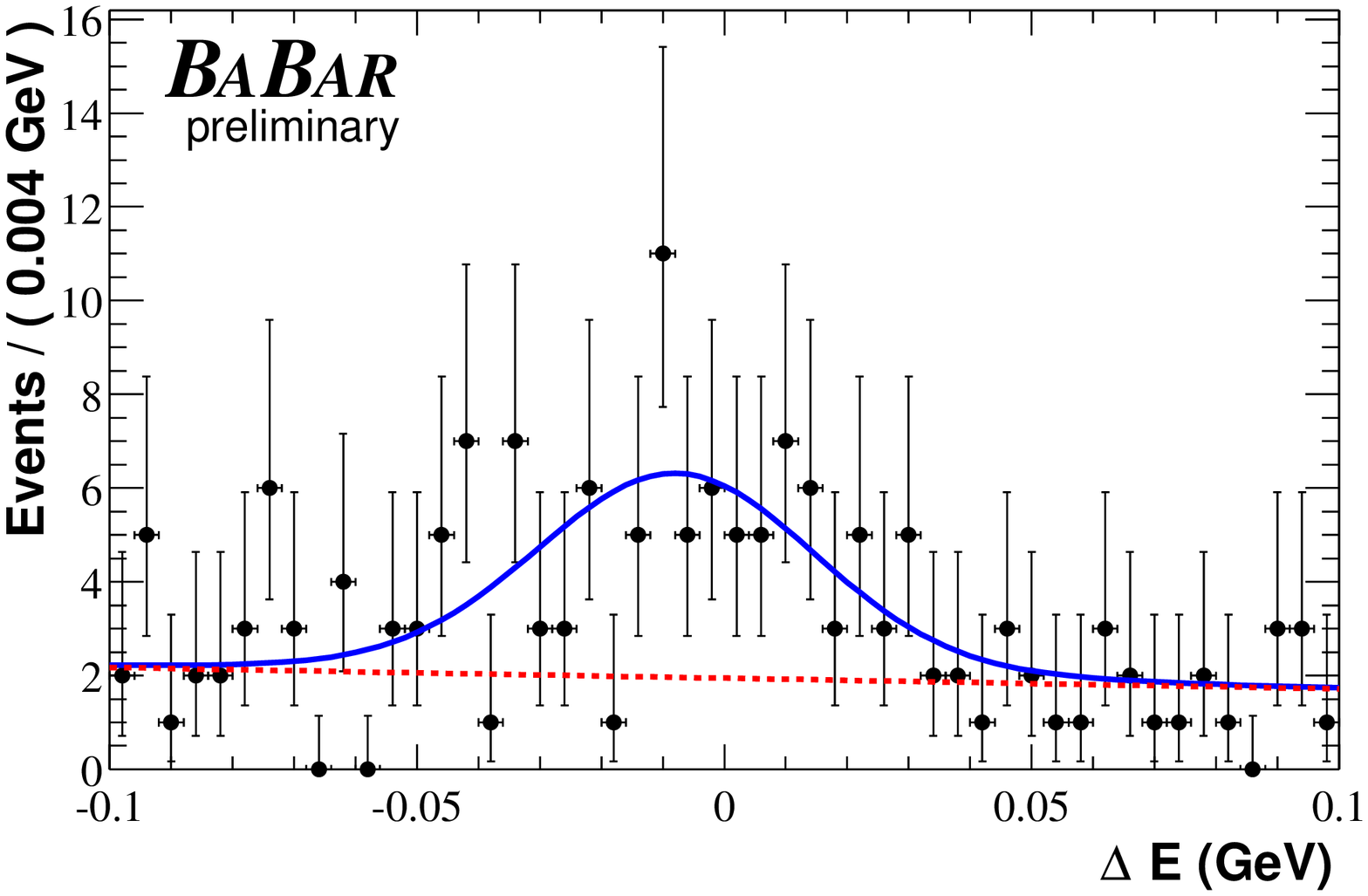} &
    \includegraphics{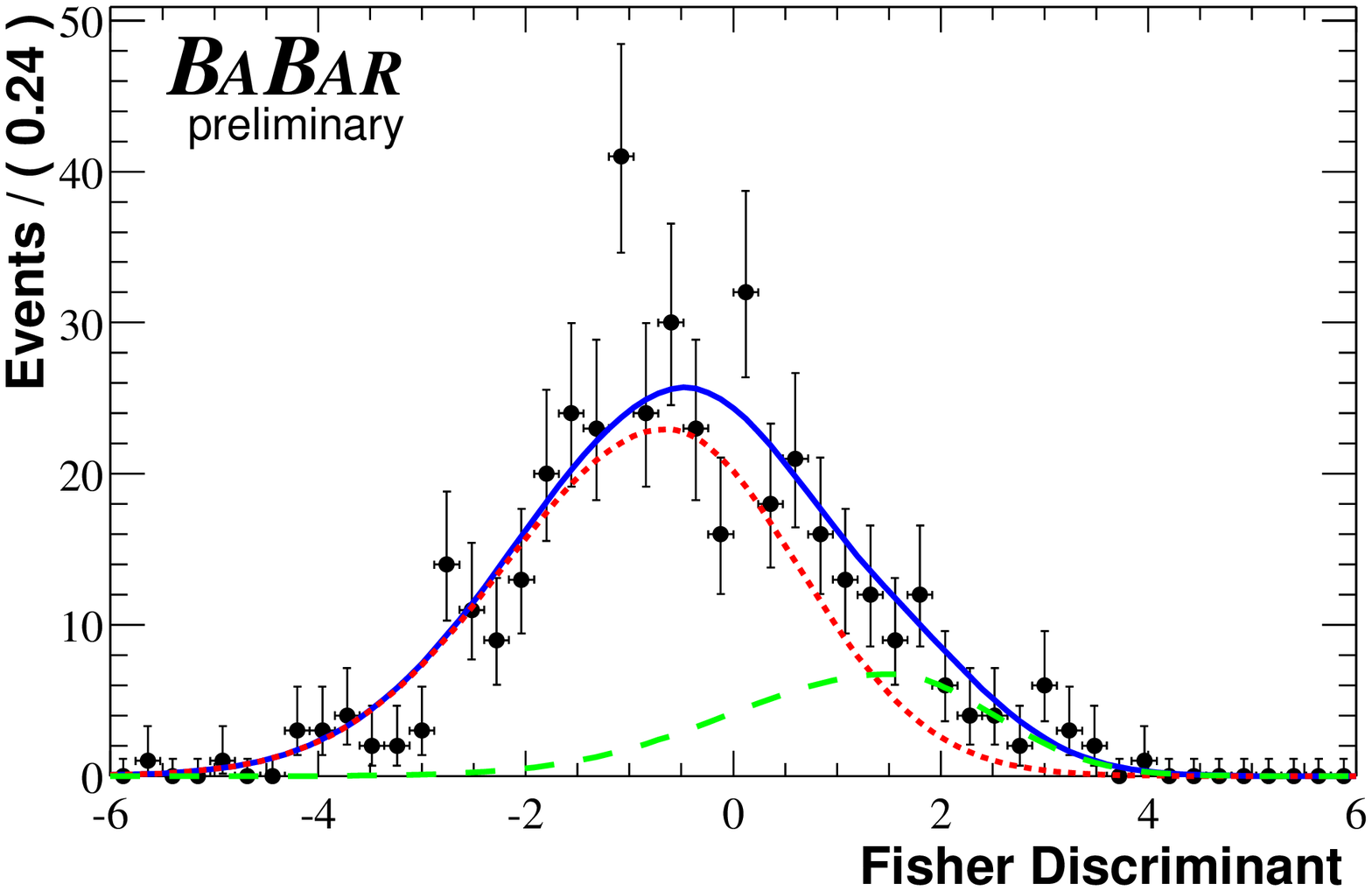} \\
    \includegraphics{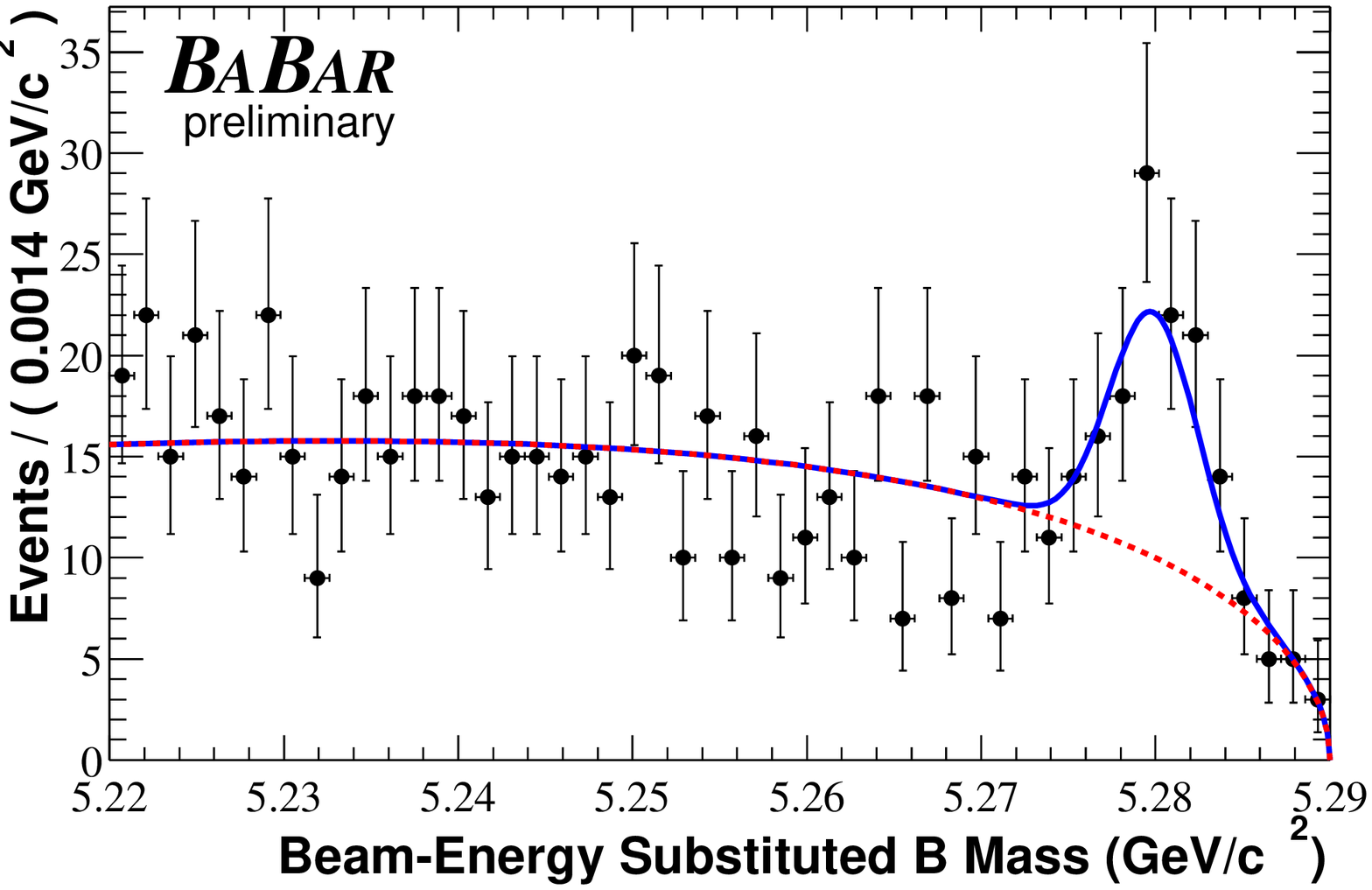} &
    \includegraphics{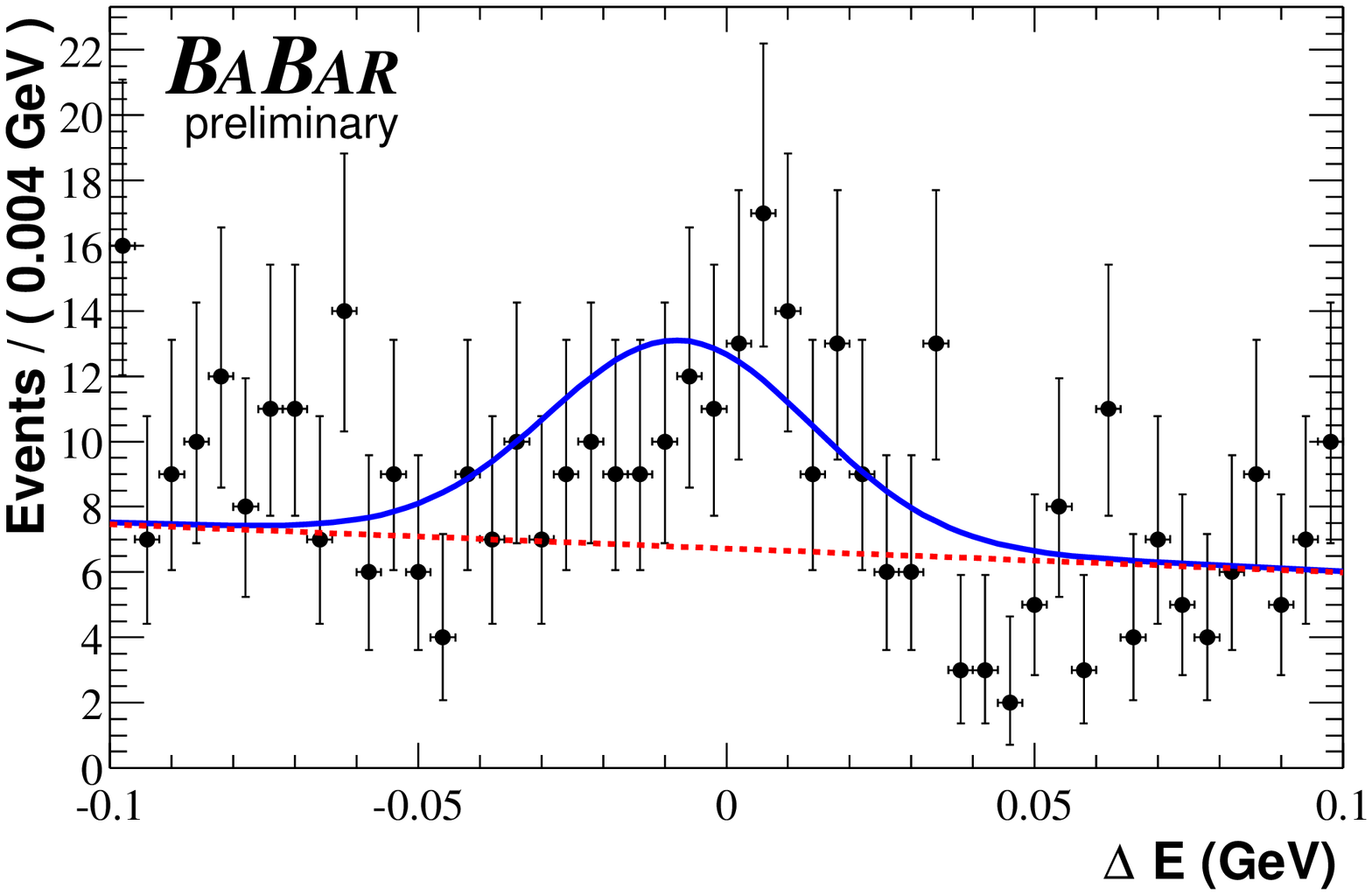} &
    \includegraphics{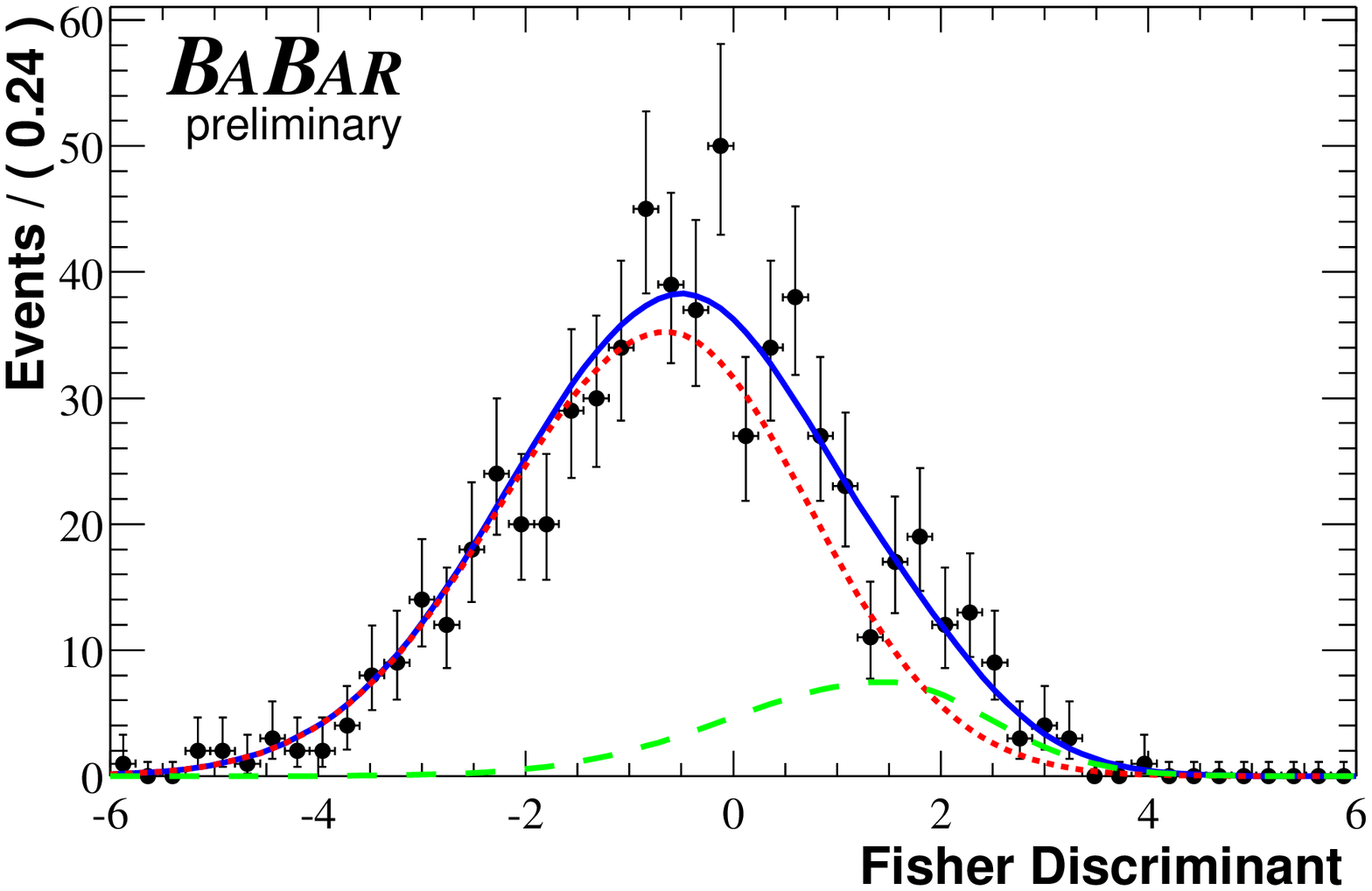} \\
    \includegraphics{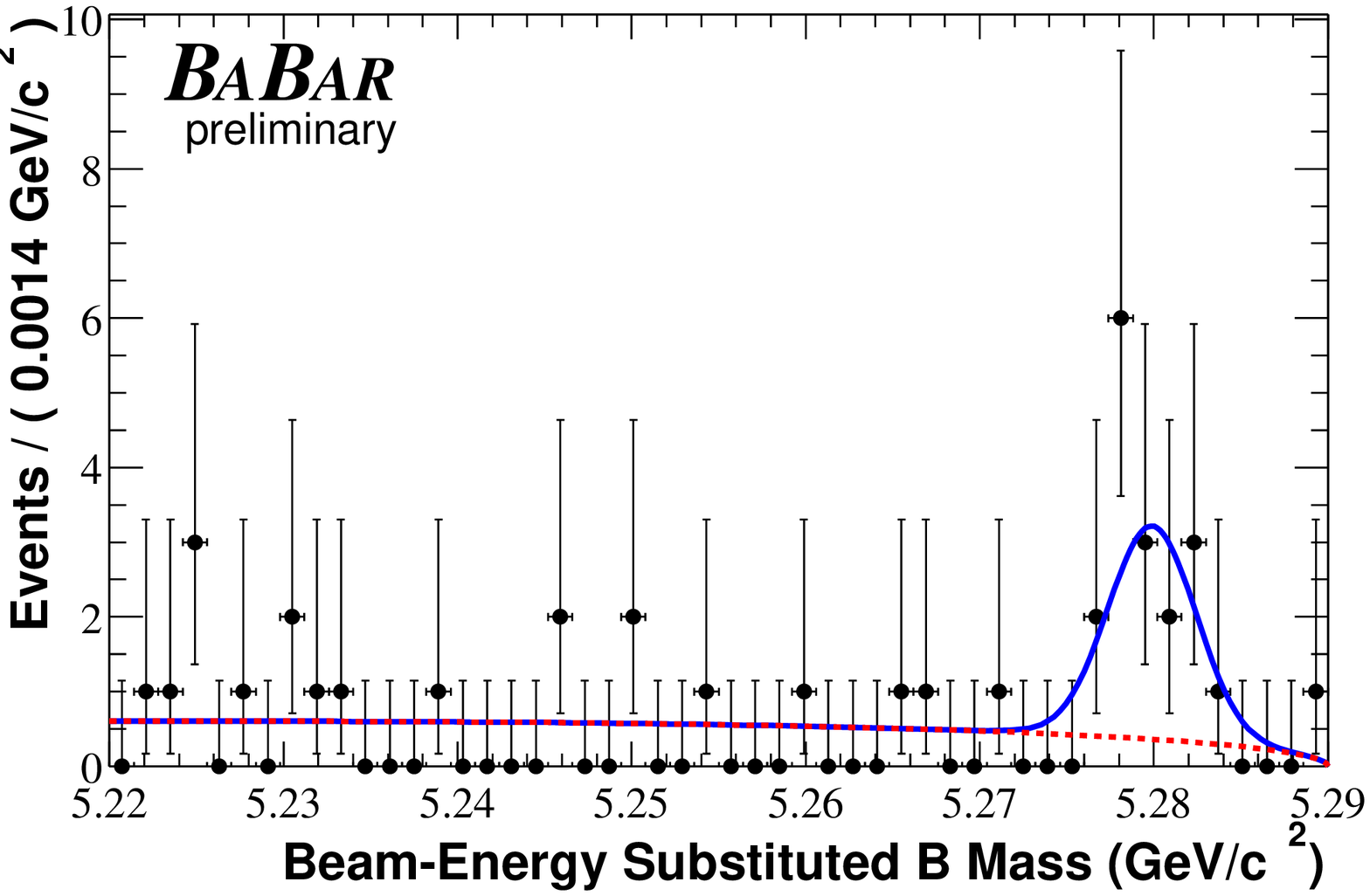} &
    \includegraphics{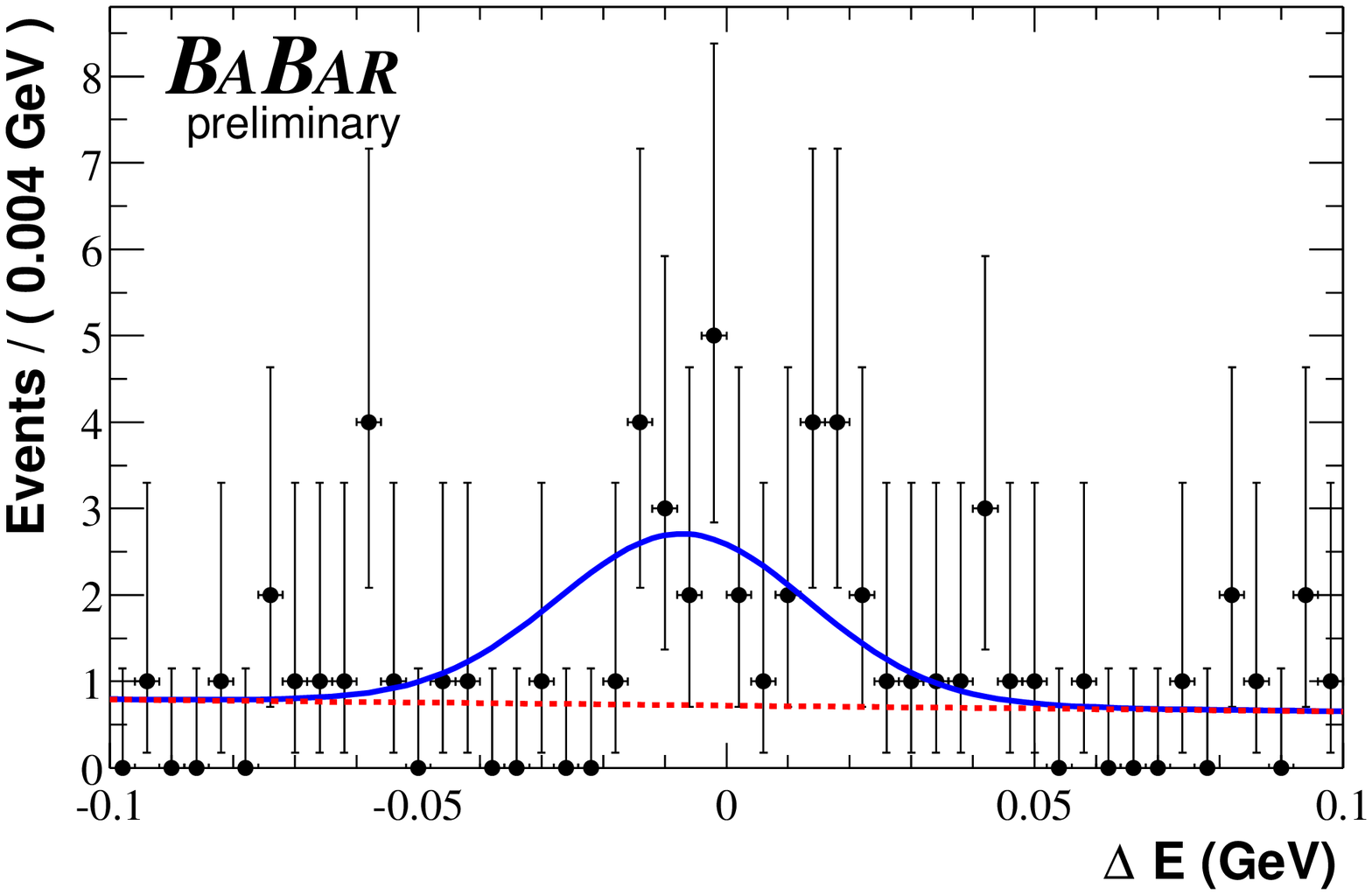} &
    \includegraphics{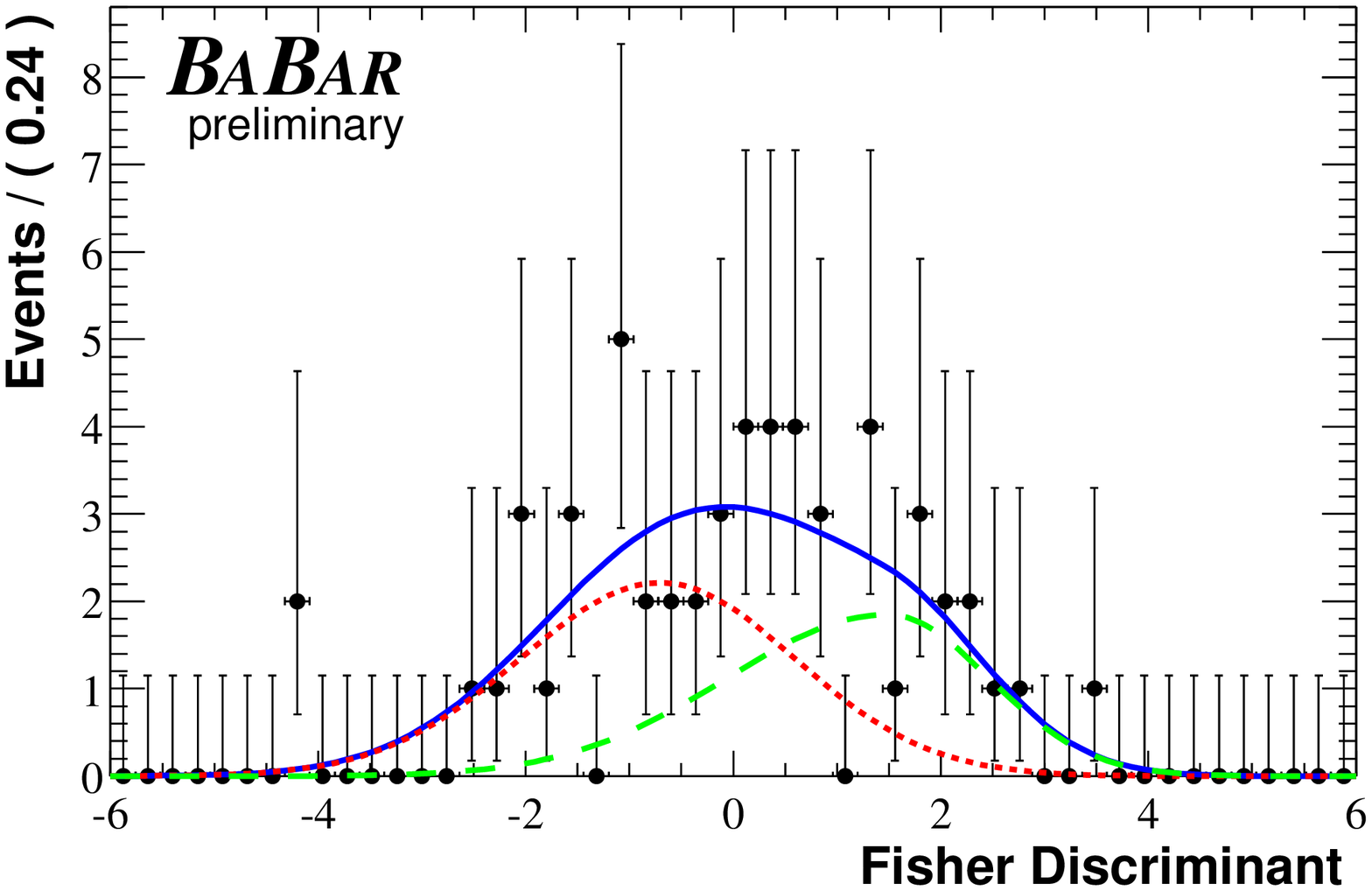}
\end{tabular}
}
\caption{Projection plots in \mes, \DE \ and \xf, produced by selecting on the event likelihood ratio 
  formed from the other two fit variables for (from top to bottom)
  regions V, VI, VII and VIII. The superimposed curve is a projection of
  the full fit with the background component shown as a dotted line
  and, for \xf, the signal component shown as a dashed
  line.}
\label{fig:ProjPlots2}  
\end{figure}

The projection plots of \mes, \DE\ and $\mathcal{F}$ for each region are
shown in Figures \ref{fig:ProjPlots1} and \ref{fig:ProjPlots2}. To
produce these plots the projected variable is excluded from the
likelihood functions and the ratio of the signal and background
likelihoods for each event evaluated. The histograms are produced with a
selection on this per--event likelihood ratio, where the selection value
was chosen separately for each region to best illustrate the signal
contribution. The projection of the fits onto that variable is then
superimposed.

\begin{figure}[htb]
\includegraphics[width=0.48\textwidth]{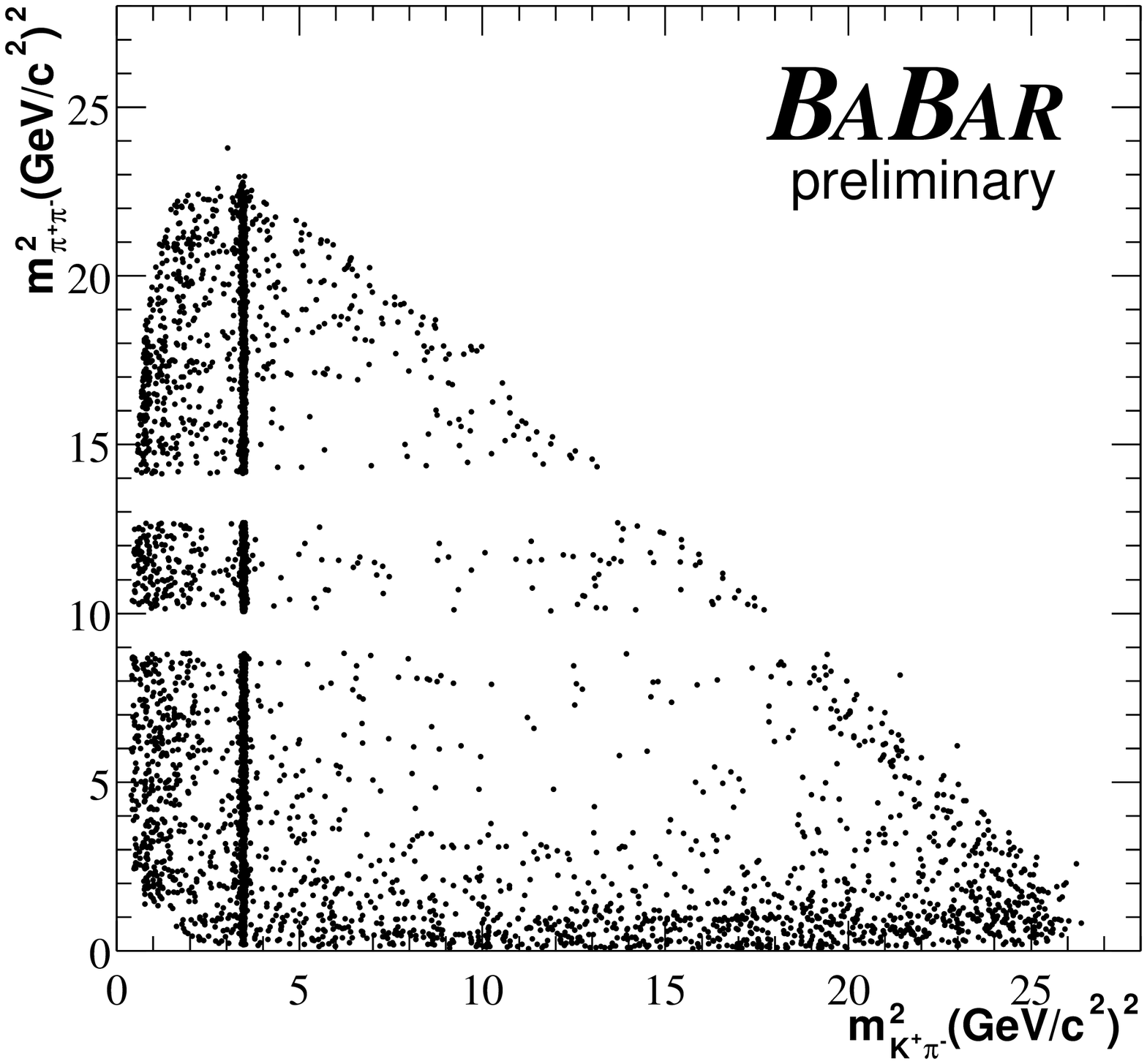}
\includegraphics[width=0.48\textwidth]{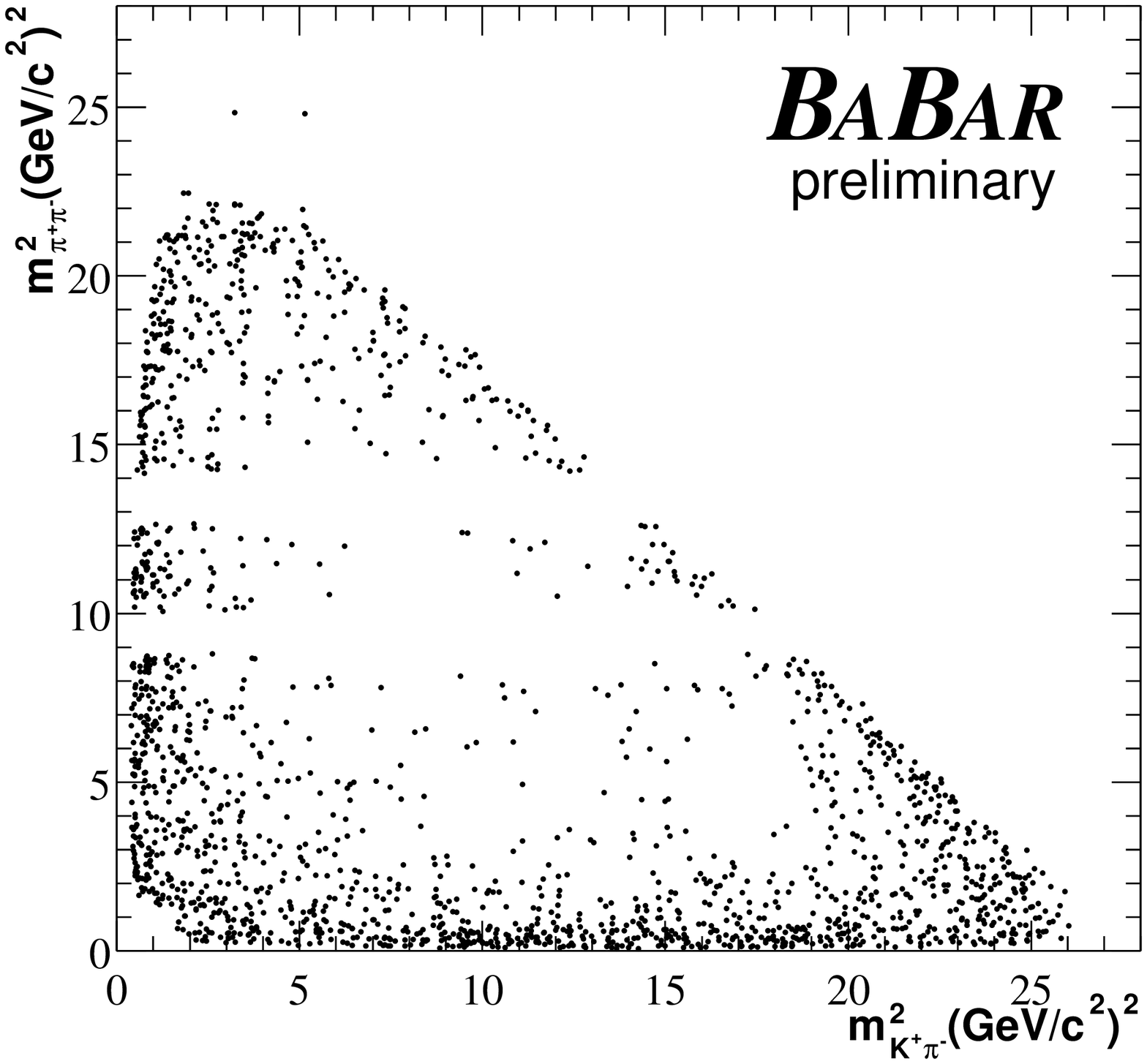}
\caption{Dalitz plots showing the observed distribution in the signal \mes \ region (left) and continuum background from the \mes sideband (right) with the mass intervals close to $J/\Psi$ and $\Psi (2S)$ removed.
}\label{fig:DalitzPlotData}
\end{figure}

Figure \ref{fig:DalitzPlotData} shows the Dalitz plot for
data events within the signal region $5.2715 < m_{ES} < 5.2865 \gevcc$ that
have a per--event likelihood ratio, formed from the \DE\ and $\mathcal{F}$ PDFs,
greater than 5. To illustrate the expected background distribution,
events passing the same likelihood selection but having a value of
$m_{ES}$ between $5.25 < m_{ES} < 5.26 \gevcc$ are shown alongside. The size of
this sideband is chosen to contain approximately the same number of
background events as are in the signal region. The mass intervals close to 
$J/\Psi$ and $\Psi (2S)$ are removed.

\begin{figure}[htb]
\resizebox{\textwidth}{!}{
    \includegraphics{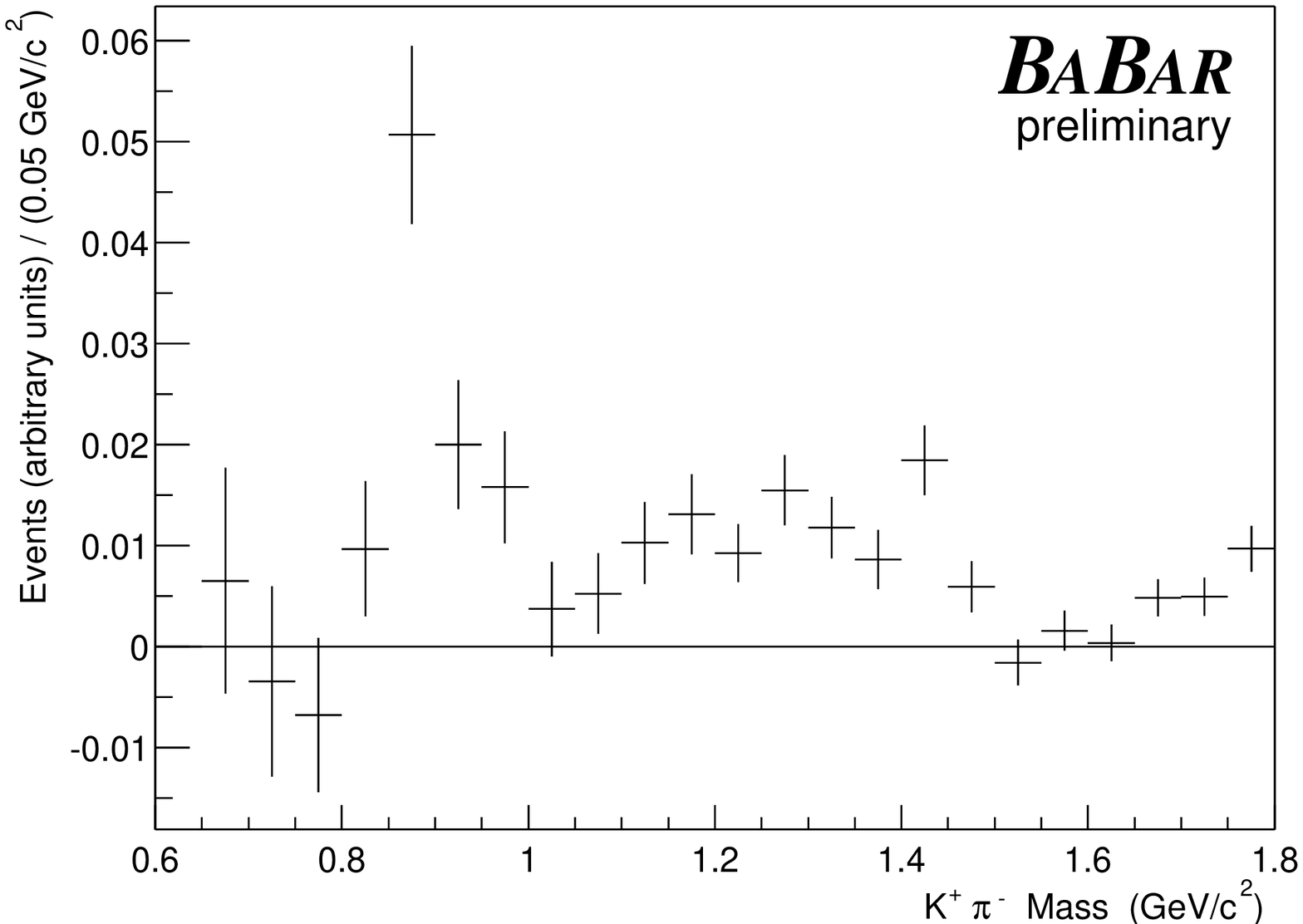}
    \includegraphics{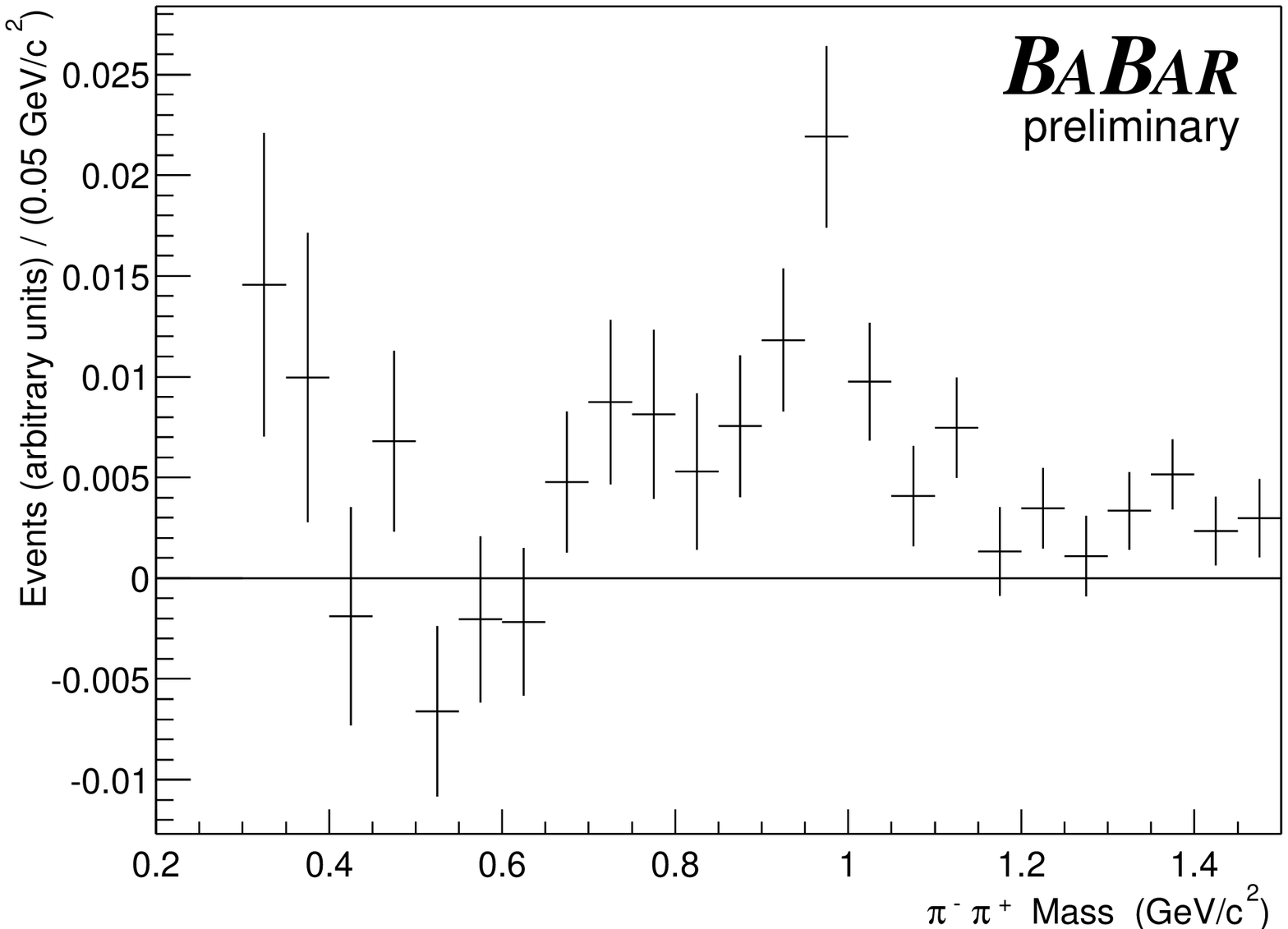}
}
\caption{Background-subtracted and efficiency-corrected projections of the Dalitz plot in $m_{K\pi}$ and $m_{\pi\pi}$ in the ranges $0.6\gevcc < m_{K\pi} < 1.8\gevcc$ and $0.2\gevcc < m_{\pi\pi} <1.5\gevcc$ with $J/\Psi$, $\Psi (2S)$ and $D^0$ vetoes applied. Peaks at the $K^{*0}(892)$ and $f_0(980)$ masses are clearly visible.}
\label{fig:kf}
\end{figure}

\begin{figure}[htbp]
\centering
\resizebox{\textwidth}{!}{
\begin{tabular}{cc}
    \includegraphics{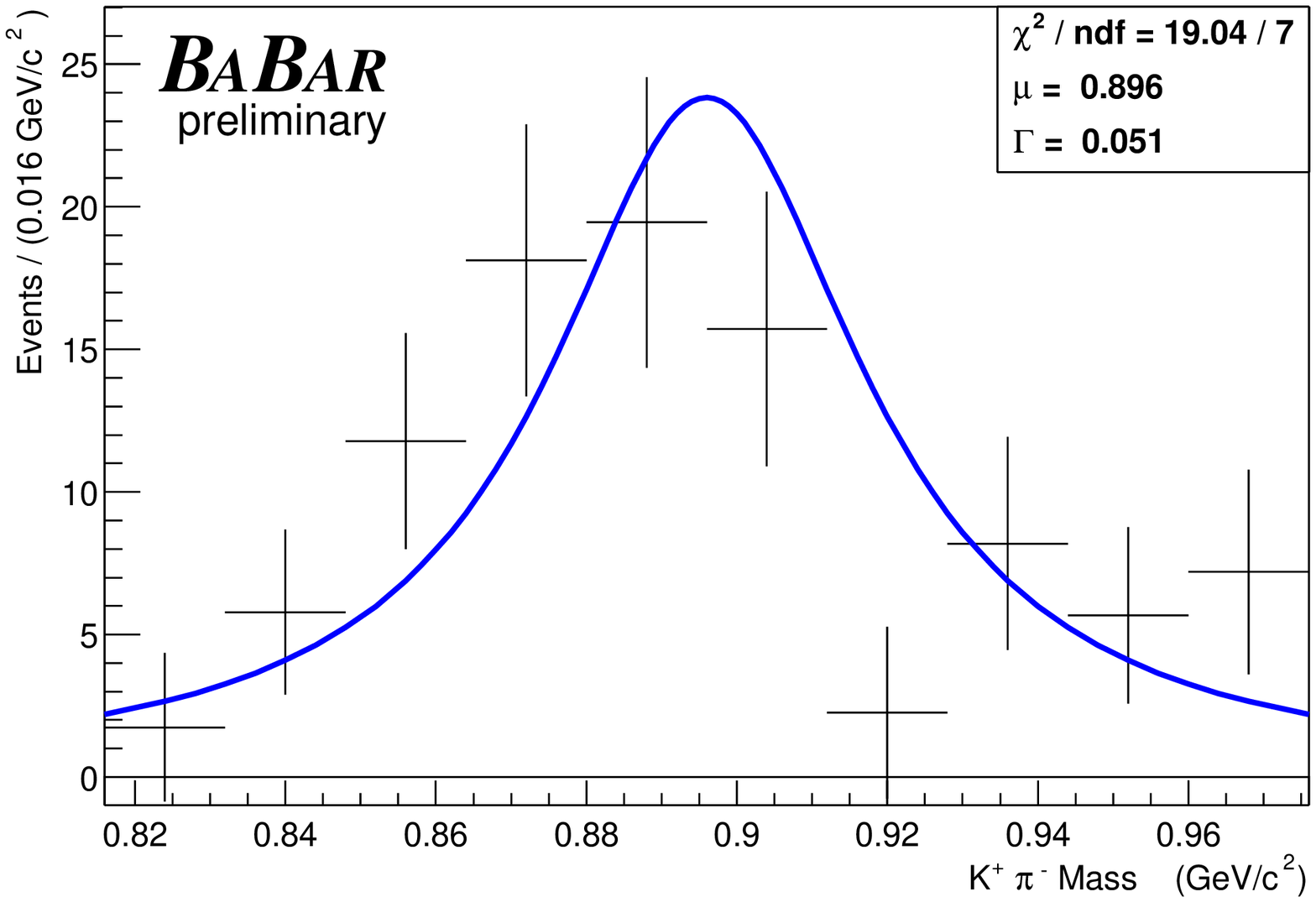} &
    \includegraphics{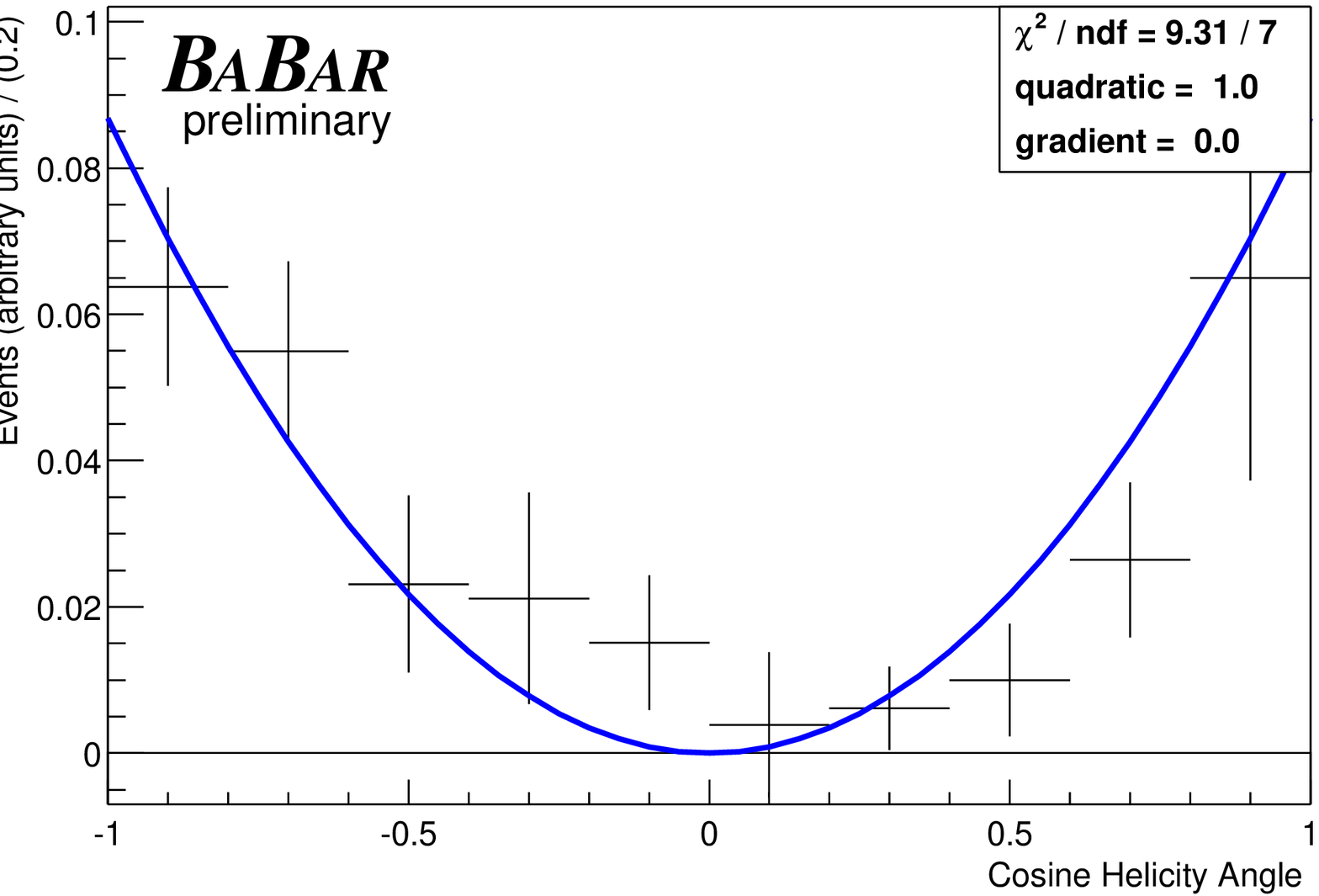} \\
    \includegraphics{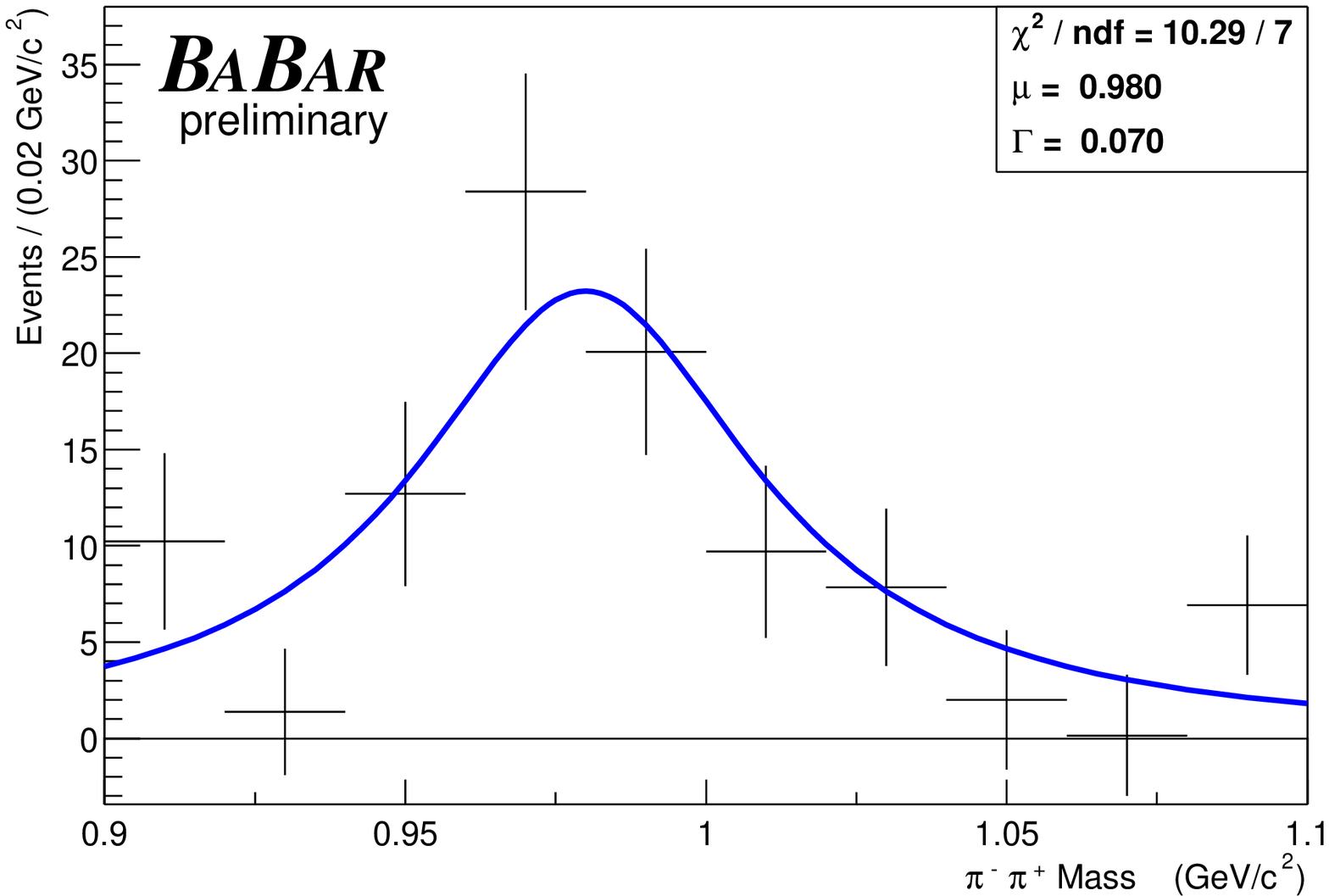} &
    \includegraphics{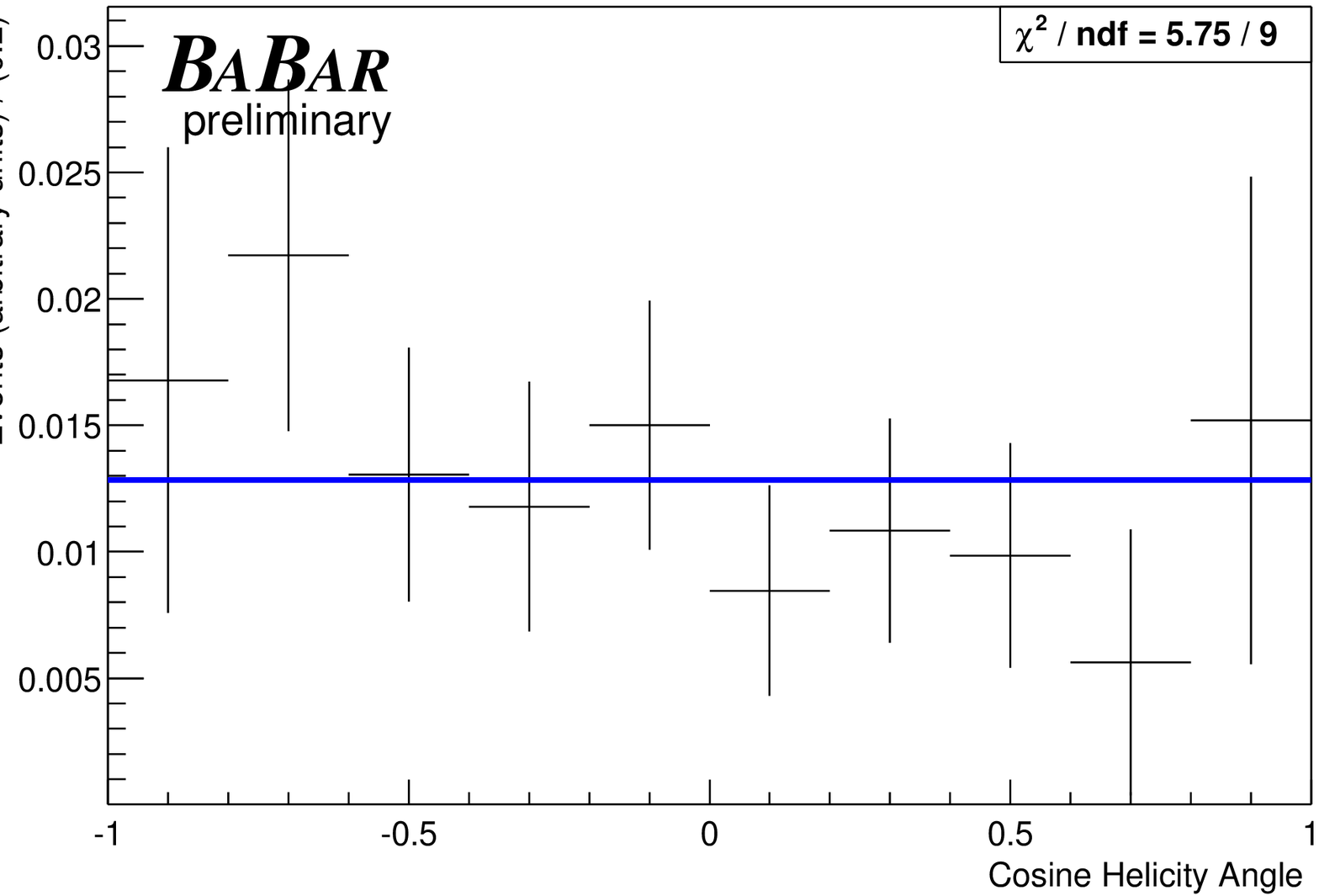} \\
\end{tabular}
}
\caption{Projection plots of the resonant mass and cosine of the helicity angle 
  for regions I and V. The background has been subtracted
  from all plots and, for the helicity distributions, a correction
  applied to each bin to account for the variation in reconstruction
  efficiency.  The bold lines are the mass and helicity angle
  distributions that would be expected from the dominant resonant mode,
  $\BpmKstarpi$ for region I and $\BpmfzK$ for region V with
  Breit-Wigner lineshapes and the masses and widths given in Table
  \ref{tab:model}.}
\label{fig:ProjPlotsHelRes}  
\end{figure}

\begin{figure}[htbp]
\centering
\resizebox{\textwidth}{!}{
\begin{tabular}{cc}
    \includegraphics{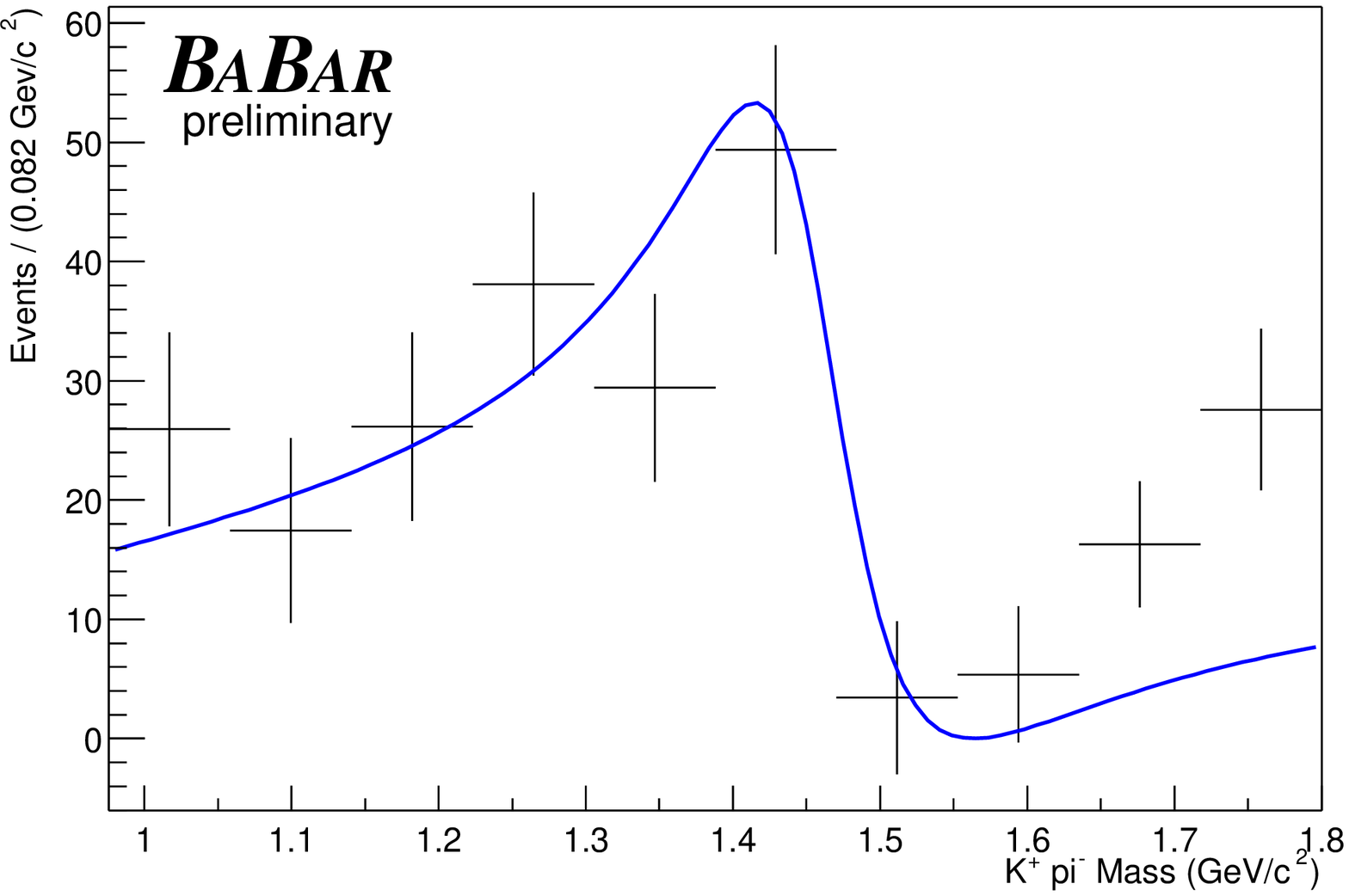} &
    \includegraphics{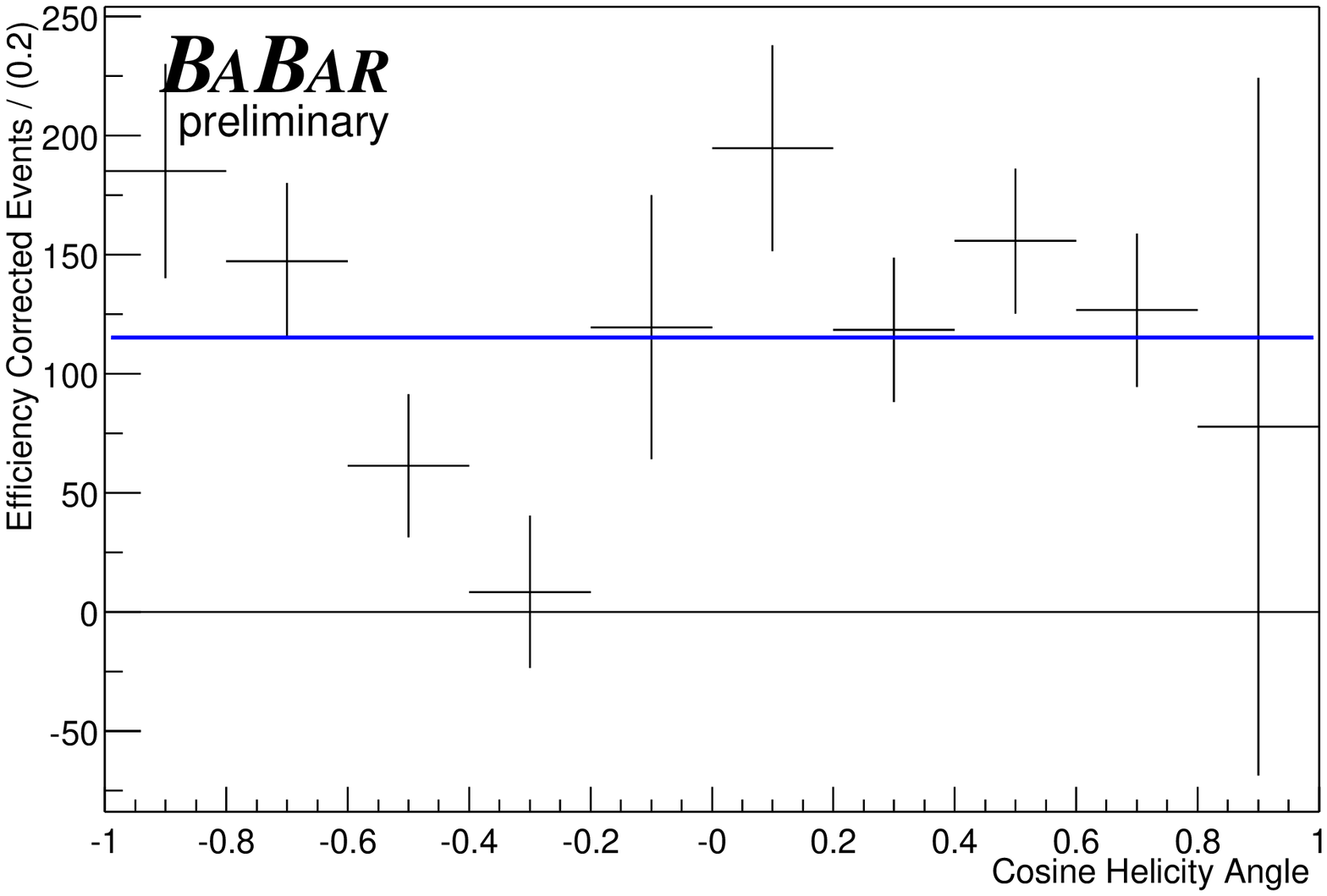} \\
\end{tabular}
} \caption{Projection plots of the resonant mass and cosine of the
  helicity angle for region II. The background has been subtracted
  from all plots and, for the helicity distributions, an efficiency
  correction applied to each bin.  The mass parameterization obtained
  from the LASS experiment is superimposed as is a flat scalar
  distribution for the cosine of the helicity angle. The increase in
  event density above 1.6 \gevcc is most likely caused by
  contributions from other higher-mass resonances.}
\label{fig:ProjPlotsHelResHighKst} \end{figure}

Figure \ref{fig:kf} shows background--subtracted projections of the
Dalitz plot in $m_{K\pi}$ from $0.6\gevcc$ to $1.8\gevcc$ and
$m_{\pi\pi}$ from $0.2\gevcc$ to $1.5\gevcc$. For these plots, signal
events are obtained using a likelihood ratio selection on the $m_{ES}$
and \xf \ PDFs and the requirement $|\DeltaE|<0.07 \gev$, while the
background is determined from the sideband $0.1 < \DeltaE< 0.35 \gev$.
The $\Dzb$, $J/\Psi$ and $\Psi (2S)$ vetoes are applied.  The plots
are efficiency--corrected using non--resonant $K^+\pi^-\pi^+$
Monte--Carlo data. Peaks at the $K^{*0}(892)$ and $f_0(980)$ masses
are clearly visible. It is not clear what other channels are
contributing, although there is a large signal in the region
$1.1<m_{K\pi}<1.4\gevcc$.  Figure \ref{fig:ProjPlotsHelRes} shows
background--subtracted plots of the resonant mass and helicity angle
distributions for regions I and V, where the $K^{*0}(892)$ and
$f_0(980)$ resonances are expected to dominate. Only the helicity
distributions have been efficiency--corrected as the efficiency does
not vary over the range of the plotted mass distributions. The plots
in Figure \ref{fig:ProjPlotsHelRes} have been overlaid with the
distribution of the expected dominant resonance using Breit--Wigner
line--shapes for the mass distributions, $\cos^2\theta_H$ for the
$K^{*0}(892)$ angular distribution and a flat line for the scalar
$f_0(980)$ angular distribution.  There is good agreement between the
overlaid and observed distributions indicating that the expected
resonances are indeed dominant in these regions.

Figure \ref{fig:ProjPlotsHelResHighKst} shows plots for region II
produced in the same way. It is not clear from these distributions which
resonances are present but it is unlikely that a single resonance can 
describe the $m_{K\pi}$ distribution.
Superimposed is the effective--range parameterization used
for the scalar $K^{*0}_0(1430)$ observed in the LASS experiment
\cite{LASS}, using parameters taken from \cite{Abele}.  The agreement
between this parameterisation and our observed distribution is good up
to 1.6 \gevcc.  The increase in event density above 1.6 \gevcc is most
likely caused by contributions from other higher resonances. The
possibility that the \Dzb is contributing to this plot has been excluded
by MC studies.

\subsection{Branching Fractions}
\label{sec:BFs}

Regions III and VIII have negligible contributions from $B^+\ra
K^+\pi^-\pi^+$ channels other than \Dzpi\ and \ChiczK, respectively.
\Dzpi\ and \ChiczK\ do not contribute to other regions.  
The branching fractions for these channels are defined by:

\begin{equation}
\label{eq:simpleBF}
  {\cal B} = \frac{\mbox{Y}}{\nbb \ \epsilon},
\end{equation}
where Y is the signal yield and $\nbb = (61.6\pm 0.7)\times 10^6$ is the number
of \BB\ events in the sample. 
It is assumed the $\FourS$ decays equally to neutral and charged $B$ meson pairs.
The reconstruction efficiency,
$\epsilon$, is calculated using signal MC and is corrected for MC/data
discrepancies in tracking and particle identification. We achieve
efficiencies of $0.330 \pm 0.017$ and $0.288 \pm 0.017$ for \Dzpi\ and
\ChiczK, respectively, where the uncertainty is purely systematic.

For the other regions, we calculate the branching fractions from the
measured yields taking into account the resonance cross--feed and
model dependence. 
$\mbox{Y}$ becomes a vector of the yields in each Dalitz region,
$\cal{B}$ becomes a vector of the branching fractions and the 
efficiency becomes a matrix $\mbox{M}$ the elements of which are the
probability of an event of a particular decay to be found in a 
particular region.

\begin{equation}
\label{eq:yield}   {\cal B} = \mbox{M}^{-1} \ \mbox{Y}/ \nbb
\end{equation}
\begin{table}
\caption{A summary of the model used to calculate branching fractions.}
\resizebox{\textwidth}{!}{
\begin{tabular} {|l|l||c|c|c|c|c|}
\hline
 $i$ & Decay Mode & Lineshape & Mass (\mevcc) & Width (\mevcc) & secondary BF (\%) & Alternative Resonance\\ 
\hline \hline
1&  \Kstarpi  & BW & $896.10\pm 0.27$ & $50.7\pm0.6$ & 33.3 & -\\
2&  $K^{*0}_0(1430) \pi^+$ & BW (LASS\cite{LASS}) & $1412\pm6$ & $294\pm23$ & $93\pm10$ & $K^{*0}_2(1430)$,$K^{*0}_1(1680)$\\
4&  \RhoK     & Blatt-W & $769.0 \pm 0.9$ & $150.9\pm1.7$ & 100 &- \\
5&  \fzK      & BW (Flatt\'e\cite{FLATTE}) & $980 \pm 10$ & $70 \pm 30$ & dominant & - \\
6&  \fzIIK    & BW & $1275\pm12$ & $185\pm30$ & $85\pm 2$ & $f_0(1370)$, $f_2(1430)$\\
7&   non-resonant     & flat & all masses & - & - & -\\         
\hline 
\end{tabular}}
\label{tab:model}
\end{table}

The branching fractions measured depend on the model of resonances assumed in calculating the efficiency matrix $\mbox{M}$.
We split $\mbox{M}$ into two component matrices, 
$\mbox{P}$ and $\epsilon$, such that each element 
$\mbox{M}_{ij} = \mbox{P}_{ij} \epsilon_{ij}$.
The $\mbox{P}$ matrix contains the event distribution around the Dalitz 
plot and the $\epsilon$ matrix contains the reconstruction efficiencies,
so the model dependence is contained in the $\mbox{P}$ matrix.

We assume one dominant contribution per region and the decay modes in
the chosen resonance model are given in Table~\ref{tab:model}.  The
\Kstarpi\ and \fzK\ channels have been seen and there is evidence for
\RhoK \cite{BelleNew,KstPi}.  For regions II, VI, VII, there are a
number of possible contributions: for our model, we choose
$K^{*0}_0(1430)$, \fzII\ and a flat non--resonant \ppK.  The masses and
widths are taken from the Review of Particle Physics \cite{PDG}, and non--relativistic Breit--Wigner line--shapes are used for all channels except for the
broad $\rho (770)$ resonance, where we use a relativistic Breit--Wigner lineshape 
with Blatt--Weisskopf damping \cite{BlattW}.  The matrix element $\mbox{P}_{ij}$ 
gives the
probability of an event of decay mode $i$ to be produced in region $j$
calculated using this model and including angular distributions and
phase space. The elements are shown in Table~\ref{tab:Pmatrix}. There
are large uncertainties in this model: the dominant resonance is unknown
in some regions, and there are uncertainties on the masses and widths of
the resonances, as well as the choice of line--shapes.  Alternative
resonances, line--shapes and the uncertainties on resonance parameters are
listed in Table~\ref{tab:model}.  These cause uncertainties in
$\mbox{P}$ and therefore on the branching fractions.  ``Model''
uncertainties on the branching fractions are evaluated that take into
consideration all of these uncertainties in the model.  There are also
uncertainties in $\mbox{P}$ due to the interference between the
resonances and these ``interference'' uncertainties are also evaluated.

\begin{table}
\caption{The elements of the matrix $\mbox{P}$ used to calculate the branching fraction central values, where $\mbox{P}_{ij}$ is the probability for an event of decay mode $i$ to be produced in region $j$. }
\begin{center}
\begin{tabular} {|l||c|c|c|c|c|c|}
\hline
                  & \multicolumn {6} {c|} {Region $j$} \\ \cline{2-7}
Decay Mode $i$       & I  & II  & IV & V & VI & VII \\ \hline  \hline

1 \Kstarpi   & 0.659 & 0.099 & 0.000 & 0.000 & 0.144 & 0.000 \\ 
2 \KstarIIpi & 0.013 & 0.720 & 0.019 & 0.018 & 0.047 & 0.110 \\ 
4 \RhoK      & 0.000 & 0.000 & 0.738 & 0.120 & 0.052 & 0.000 \\ 
5 \fzK       & 0.000 & 0.000 & 0.084 & 0.779 & 0.084 & 0.000 \\ 
6 \fzIIK     & 0.008 & 0.031 & 0.020 & 0.064 & 0.741 & 0.111 \\
7 high mass  & 0.013 & 0.127 & 0.032 & 0.029 & 0.074 & 0.642 \\ 
\hline
\end{tabular}
\label{tab:Pmatrix}
\end{center}
\end{table}

The element $\epsilon_{ij}$ is defined as the number of decay--mode $i$
MC events with a candidate in region $j$ passing all the selection
criteria divided by the number of decay mode $i$ events generated in
region $j$.  The $\epsilon$ matrix is calculated using resonant signal
MC for $\BpmKstarpi$, $\BpmRhoK$ and $\BpmfzK$, and non--resonant MC for
other decay modes.  These efficiencies have negligible dependence on the
resonance line--shape; however, they are affected by the angular
distribution.  Corrections for MC/data discrepancies in particle
identification and tracking are applied.  The efficiencies are shown in
Table~\ref{tab:epsilon}.

\begin{table}
\caption{The elements of the efficiency matrix $\epsilon$, where 
$\epsilon_{ij}$ is the efficiency of decay mode $i$ events
produced in region $j$ to pass the selection cuts. The statistical uncertainty
on these numbers is negligible and the fractional systematic uncertainty is $5.8\%$ (see text for details).} 
\begin{center}
\begin{tabular} {|l||c|c|c|c|c|c|}
\hline
                    & \multicolumn {6} {c|} {Region $j$} \\ \cline{2-7}
 Decay Mode $i$       & I & II  & IV & V & VI & VII  \\
\hline \hline
1 \Kstarpi   &0.364   &0.360   &-      &-      &0.169  &-   \\
2 scalar $K^{*0}$ &0.352   &0.329   &0.357  &0.353  &0.372  &0.334  \\ 
4 \RhoK      &-       &-       &0.287  &0.273  &0.286  &-     \\
5 \fzK       &-       &-       &-      &0.349  &0.362  &-     \\
6 scalar $f$ &0.352   &0.329   &0.357  &0.353  &0.372  &0.334  \\ 
7 high mass  &0.352   &0.329   &0.357  &0.353  &0.372  &0.334  \\ 
\hline 
\end{tabular}
\label{tab:epsilon}
\end{center}
\end{table}

\begin{table}
\caption{The branching fractions and uncertainties for the channels measured.}
\begin{center}
\begin{tabular}{|l||c|c|c|c|c|}
\hline
Channel & BF & \multicolumn {4} {c|} {uncertainties ($\times 10^{-6}$)} \\ \cline{3-6}
        & ($\times 10^{-6}$)   & stat & sys & model & interference \\
    \hline\hline
 $K^{*0}(892)\pi^+, K^{*0}\ra K^+\pi^-$    & 10.3 & $\pm 1.2$ & $\pm 0.7$ & $^{+0.4}_{-2.5}$ & $\pm 0.6$  \\ \hline
``higher $K^{*0}$''$\pi^+,K^{*0}\ra K^+\pi^-$   & 25.1 & $\pm 2.0$ & $\pm 2.9$ & $^{+9.4}_{-0.5}$ & $\pm 4.9$  \\ \hline
 \Dzb $\pi^+, \Dzb \to K^+ \pi^-$       & 184.6& $\pm 3.2$ & $\pm 9.7$ & - & - \\\hline
 $\rho^0(770)K^+,\rho^0(770)\ra\pi^+\pi^-$ & 3.9  & $\pm 1.2$ & $^{+0.3}_{-0.6}$ &$^{+0.3}_{-3.2}$ & $\pm 1.2$  \\ \hline
 $f_0(980)K^+, f_0\ra\pi^+\pi^-$           & 9.2  & $\pm 1.2$ & $\pm 0.6$ & $^{+1.2}_{-1.9}$ & $\pm 1.6$ \\ \hline
``higher $f$''$K^+, f\ra\pi^+\pi^-$            & 3.2  & $\pm 1.2$ & $\pm 0.5$ & $^{+5.8}_{-2.4}$ & $\pm 1.5$  \\ \hline
 Non-resonant                              & 5.2  & $\pm 1.9$ & $^{+0.8}_{-1.8}$ & $^{+3.3}_{-7.5}$ & $\pm 6.4$  \\ \hline     
 $\chi_{c0} K^+, \chi_{c0}\to\pi^+\pi^-$& 1.5  & $\pm 0.4$ & $\pm 0.1$ & - & - \\
\hline	
\end{tabular}   
\label{table:BFs}
\end{center}
\end{table}

Table~\ref{table:BFs} gives the branching fractions produced from the
vector of yields with statistical, reconstruction--systematic, 
resonance--model and interference uncertainties.

\subsubsection{Systematic Uncertainties}

In most regions, the largest contribution to the reconstruction
systematic uncertainty, between 3\% and 11\%, is from the
uncertainties on the PDF parameters.  The $B$ background subtraction
uncertainty is below 6\% for all channels except for \BpmRhoK\ (18\%)
and non--resonant (32\%), where it is dominant.  The systematic
uncertainties from track efficiency corrections (2.4\%), particle
identification efficiency (4.5\%), $\cos\theta_T$ (2.8\%) and the number
of \BB\ pairs (1.1\%) are independent of the region considered. For
``higher $f$'' and ``higher $K^{*0}$'', there is an additional
contribution of 9\% to account for the selection efficiency being
measured using scalar MC though the true angular distribution of the
contribution is unknown.

The model uncertainties include the effect of uncertainties in the
resonance contributions, resonance masses, widths and line--shapes of
the model, given in Table~\ref{tab:model}.  The possibility that the
dominant contribution to a region may be from another resonance is
taken into account for $K^{*0}_0(1430)$ and $f_2(1270)$, where the
alternative resonances are listed in Table~\ref{tab:model}.  We
consider the possibility that the component measured in Region VII
does not extend into the other regions. We also allow for the possibility
that the $f_0(980)$ resonance is described by the Flatt\'e line--shape
\cite{FLATTE} and the $K^{*0}_0(1430)$ has the line--shape suggested by
data from the LASS experiment \cite{LASS}\cite{Abele} instead of the
Breit--Wigner form.  We estimate model uncertainties on the branching
fractions by varying the model contributions, resonance parameters and
line--shapes within the parameters given in Table~\ref{tab:model},
calculating a new $P$ matrix and recalculating the branching
fractions.  The model uncertainty is the quadratic sum of all the
variations in the branching fraction due to these individual changes.

The effect of interference on the branching fractions is evaluated
by generating many Dalitz plots with the observed branching fractions 
but with each contribution having a random phase and allowing 
interference, then measuring the branching fractions using the 
$\mbox{P}$ matrix in the same way as done on the data. 
This produces a range of branching fractions for each channel with the
correct branching fraction as the mean.  The RMS variation is
taken to be the uncertainty due to interference and this is shown 
in Table~\ref{table:BFs} under the column ``interference''.

\subsection{Conclusion}

In conclusion, we have made preliminary measurements of the branching fractions for the following
channels with a statistical significance greater than 5$\sigma$. This
significance for a particular branching fraction ${\cal B}_a$ is
evaluated by assuming that branching fraction is zero and minimizing, as
a function of the other branching fractions, the $\chi^2$ separation of the
measured yields and those obtainable from the branching fractions
themselves.  The significance is then given by $\sqrt{\chi^2({\cal B}_a=0)}$.

\begin{itemize}
\item $\calB(\BpmKstarpi, K^{*0}\ra K^+\pi^-) = \BrKstarpiVal$,
\item $\calB(\BpmfzK, f_0\ra \pi^+\pi^-) = \BrfzKVal$, 
\item $\calB(\BpmChiczK, \chi_{c0}\ra \pi^+\pi^-) = \BrChiczKVal$,
\item $\calB(\BpmDzpi, \Dzb \ra K^+\pi^-) = \BrDzpiVal$ and
\item $\calB(B^+\ra$``higher $K^{*0}$''$\pi^+) = \BrKstarIIpiVal$, where ``higher $K^{*0}$'' means any combination of $K^{*0}_0(1430), K^{*0}_2(1430)$ and $K^{*0}_1(1680)$.
\end{itemize}

\noindent The first uncertainty is statistical and the second is systematic.
This analysis has taken into account the uncertainty in the knowledge
of the nature and parameterization of these resonances as well as
interference between them, and these uncertainties are included (added
in quadrature) in the above systematic value.  Using the value of
${\cal B}(K^{*0}(892)\ra K^+\pi^-)=2/3$, we find $\calB(\BpmKstarpi) =
(15.5 \pm 1.8 ^{+1.5}_{-3.2})\times 10^{-6}$, which is consistent
with, and more precise than, previous
measurements\cite{BelleNew,KstPi}.  The observation of the decay
\BpmfzK \ is statistically significant, 
providing hints about the $f_0(980)$ production mechanism in
the $B$ system. There is also a significant signal for $B^{+} \to$
``higher $K^{*0}$'' $\pi^{+}$, which has a mass distribution which is
partly described by the $K_0^{*0}(1430)$, as observed by the LASS
experiment.

We give 90\% confidence-level upper limits for the branching fractions of the
following channels, including the non--resonant
component. This upper limit is taken as the value of that branching
fraction for which the minimum $\chi^2$ separation of the measured
yields and those calculated from the branching fractions is 1.64.

\begin{itemize}
\item $\calB(\BpmRhoK) < 6.2\times 10^{-6}$,
\item $\calB(B^+ \ra K^+\pi^-\pi^+$ non--resonant$) < 17\times 10^{-6}$ and
\item $\calB(B^+\ra$``higher $f$''$ K^+) < 12 \times 10^{-6}$, where ``higher $f$'' means a combination of $f_2(1270)$, $f_0(1370)$ and $f_2(1430)$. 
\end{itemize}

\noindent The tight limit on the non--resonant component means that it will be difficult
to obtain the Unitarity Triangle angle $\gamma$ by the methods of Refs.
\cite{gamma1} and \cite{ref:GammaPaper}.

 \section{Acknowledgements}

We are grateful for the 
extraordinary contributions of our \pep2\ colleagues in
achieving the excellent luminosity and machine conditions
that have made this work possible.
The success of this project also relies critically on the 
expertise and dedication of the computing organizations that 
support \babar.
The collaborating institutions wish to thank 
SLAC for its support and the kind hospitality extended to them. 
This work is supported by the
US Department of Energy
and National Science Foundation, the
Natural Sciences and Engineering Research Council (Canada),
Institute of High Energy Physics (China), the
Commissariat \`a l'Energie Atomique and
Institut National de Physique Nucl\'eaire et de Physique des Particules
(France), the
Bundesministerium f\"ur Bildung und Forschung and
Deutsche Forschungsgemeinschaft
(Germany), the
Istituto Nazionale di Fisica Nucleare (Italy),
the Foundation for Fundamental Research on Matter (The Netherlands),
the Research Council of Norway, the
Ministry of Science and Technology of the Russian Federation, and the
Particle Physics and Astronomy Research Council (United Kingdom). 
Individuals have received support from 
the A. P. Sloan Foundation, 
the Research Corporation,
and the Alexander von Humboldt Foundation.

\end{document}